\documentclass[12pt,showpacs,preprint,nofootinbib,showkeys]{revtex4}

\usepackage[english]{babel}
\usepackage{bbm}
\usepackage[vcentermath]{youngtab}
\usepackage{amssymb}
\usepackage{epic}
\usepackage{eepic}

\font\af=msbm10

\newcommand{\cD}[2]{\mathcal{D}^{\,#1}_{#2}}
\newcommand{\unitop}{\mathbbm{1}}
\newcommand{\ud}{\mbox{d}}
\newcommand{\im}{\mbox{i}}
\newcommand{\ex}{\mbox{e}}
\newcommand{\eqnn}[1]{\begin{eqnarray*}#1\qquad
\quad\qquad\end{eqnarray*}}
\newcommand{\eqnl}[2]{\par\parbox{14cm}{\begin{eqnarray*}#1\end{eqnarray*}}\hfill\parbox{1cm}{\begin{eqnarray}\label{#2}\end{eqnarray}}\break}

\newcommand{\eqn}[1]{\begin{eqnarray*}#1\end{eqnarray*}}
\newcommand{\eqngr}[2]{\begin{eqnarray*}#1\\#2\end{eqnarray*}}
\newcommand{\eqngrlb}[3]{\par\parbox{11cm}{\begin{eqnarray}\fbox{$\displaystyle#1\\#2$}\end{eqnarray}}\hfill\parbox{1cm}{\begin{eqnarray}\label{#3}\end{\eqnarray}}\break}
\newcommand{\eqngrl}[3]{\par\parbox{14cm}{\begin{eqnarray*}#1\\#2\end{eqnarray*}}\hfill\parbox{1cm}{\begin{eqnarray}\label{#3}\end{eqnarray}}\break}

\newcommand{\refs}[1]{(\ref{#1})}

\def\cT{\mathcal T}
\def\cY{\mathcal Y}
\def\cN{{\mathcal N}}
\def\cC{{\cal C}}
\def\cH{{\cal H}}

\def\C{\mbox{\af C}}
\def\Z{\mbox{\af Z}}
\def\eps{\epsilon}
\def\br{\textbf{r}}
\def\bp{\textbf{p}}

\def\bL{\textbf{L}}
\def\bC{\textbf{C}}
\def\bK{\textbf{K}}
\def\R{\mbox{\af R}}
\def\id{\mathbbm{1}}
\def\tr{\textrm{tr}\;}

\def\ket#1{\mathinner{|{#1}\rangle}}
\def\ms{\!-\!}
\def\ps{\!+\!}
\def\vac{|0\rangle}
\def\psid{\psi^\dagger}
\def\psia{\psi_a}
\def\psiad{\psi_a^\dagger}
\def\psib{\psi_b}
\def\psibd{\psi_b^\dagger}
\def\ra{\rangle}
\def\la{\langle}

\def\gam{\gamma}

\def\mtxt#1{\quad\hbox{{#1}}\quad}
\def\ga{\gamma}
\def\Ga{\Gamma}

\def\al{\alpha}
\def\lam{\lambda}

\def\Qd{Q^\dagger}

\def\p{\partial}
\def\pa{\partial}

\def\ha{{1\over 2}}
\def\ov{\over}

\def\R{\mbox{\af R}}
\def\Z{\mbox{\af Z}}
\def\N{\mbox{\af N}}

\def\has{{1\over\sqrt{2}}}
\def\phid{\phi^\dagger}
\def\phib{\bar\phi}

\begin{document}

\vspace{2 cm}
\title {Algebraic Solution of the Supersymmetric Hydrogen Atom\\ 
in $d$ Dimensions}
\hfill{FSU TPI 02/02}
\author{A. Kirchberg\footnotemark[1], J.D. L\"ange\footnotemark[1], 
P.A.G. Pisani\footnotemark[2] and A. Wipf\footnotemark[1]}
\vspace{1 cm}

\affiliation{
\footnotemark[1]Theoretisch-Physikalisches Institut, Friedrich
Schiller Universit\"at Jena, Fr\"obelstieg 1, 07743 Jena,
Germany \footnotetext[1]{\tt A.Kirchberg@tpi.uni-jena.de, 
J.D.Laenge@tpi.uni-jena.de, A.Wipf@tpi.uni-jena.de} \\
\footnotemark[2]IFLP, Departamento de F\'{\i}sica de Ciencias Exactas, 
UNLP C.C. 67, 1900 La Plata, 
Argentina\footnotetext[2]{\tt Pisani@obelix.fisica.unlp.edu.ar}}

\vspace{1 cm}

\begin{abstract}
\noindent
In this paper the $\cN\!=\!2$ supersymmetric extension of the Schr\"odinger
Hamiltonian with $1/r$-potential in arbitrary
space-dimensions is constructed. The 
supersymmetric hydrogen atom admits a conserved 
Laplace-Runge-Lenz vector which extends the 
rotational symmetry $SO(d)$ to a hidden $SO(d\ps 1)$ 
symmetry. This symmetry of the system is used to determine the discrete eigenvalues with 
their degeneracies and the corresponding bound state wave functions.
\end{abstract}

\maketitle

\setlength{\parindent}{0cm}

\section{Introduction}
\label{sec1}

For a closed system of two non-relativistic point masses interacting 
via a central force the angular momentum $\bL$
of the relative motion is conserved
and the motion is always in the plane perpendicular to
$\bL$.
If the force is derived from the Newton or Coulomb potential,
there is an additional conserved quantity: the 
Laplace-Runge-Lenz\footnote{A more suitable name for
this constant of motion would be Hermann-Bernoulli-Laplace
vector, see \cite{goldstein}.} vector \cite{history}. 
For the hydrogen atom this vector has the form
\[
\bC={1\ov m}\bp\times\bL-{e^2\ov r} \br \;,
\qquad \bL=\br\times \bp \;,
\]
where $m$ denotes the reduced mass of the proton-electron
system. The Laplace-Runge-Lenz vector is perpendicular to $\bL$ and
hence is a vector in the plane of the orbit. It points in
the direction of the semi-major axis.

Quantum mechanically, 
one defines the hermitian Laplace-Runge-Lenz vector
\eqnl{
\bC={1\ov 2m}\big(\bp\times\bL-\bL\times\bp\big)-{e^2\ov r}\br \; .}{einf2}
By exploiting the existence of this conserved vector operator,
Pauli calculated the spectrum of the hydrogen atom by purely
algebraic means \cite{pauli, fock}. He noticed that 
the angular momentum $\bL$ together with the vector
\[
\bK=\sqrt{{-m\ov 2H}}\,\bC \;,\]
which is well-defined and hermitian on bound states with
negative energies,
generate a hidden $SO(4)$ symmetry algebra,
\[
[L_a,L_b]=\im\hbar \eps_{abc}L_c \;,\quad
[L_a,K_b]=\im\hbar \eps_{abc}K_c \;,\quad
[K_a,K_b]=\im\hbar \eps_{abc}L_c \;,
\]
and that the Hamiltonian
can be expressed in terms of $\bK^2+\bL^2$, one of the two second-order
Casimir operators of this algebra, as follows
\eqnl{
H=-{me^4\ov 2}{1\ov \bK^2+\bL^2+\hbar^2} \;.}{einf5}
One further observes that the other Casimir operator 
$\bK\cdot\bL$ vanishes and arrives
at the bound state energies by purely group
theoretical methods. The existence
of the conserved vector $\bK$ also explains the 
accidental degeneracy of the hydrogen spectrum.
Only much later the scattering amplitude 
of the hydrogen atom has been calculated with
the help of the Laplace-Runge-Lenz vector \cite{zwanziger}.

In this paper we shall generalize these
results in two directions: 
first to the hydrogen atom in arbitrary 
dimensions\footnote{When speaking of the $d-$dimensional
hydrogen atom, we always mean the $1/r$-potential, although
this potential permits the application of Gauss' law 
in three dimensions only.} \cite{sudarshan}
and second to the corresponding supersymmetric extensions.
In the following section we prove that the Schr\"odinger 
Hamiltonian in $d$ dimensions with $1/r$ potential
admits a generalization of the Laplace-Runge-Lenz vector. 
Together with the generators of
the rotation group $SO(d)$
this vector generates  the dynamical symmetry group  
$SO(d+1)$. This hidden symmetry allows 
then for a purely algebraic solution 
of the hydrogen atom in arbitrary dimensions,
very much as in three dimensions.

In section \ref{sec3} we summarize the extensions of
$d$-dimensional Schr\"odinger Hamiltonians to models with
$\cN=2$ supersymmetry. The corresponding Hamiltonians
may be written as $2^d\times 2^d$-dimensional matrix 
Schr\"odinger operators. For a particular choice of
the superpotential we obtain the supersymmetric
extension of the hydrogen atom. For instance in $d=3$ (and in a suitable basis) we find the Hamiltonian
\eqn{H = \left(p^2 + \lambda^2\right) \unitop_{8} - \frac {2\lambda}{r} \left(\begin{array}{cccc} 1 & & & \\ & M_1 & & \\ & &
M_2 & \\ & & & -1 \end{array} \right)\;,}
with $3\times3$ matrices
\eqn{M_{1 \; ab} =  \hat x_a \hat x_b\;, \qquad M_{2 \; ab} = \delta_{ab} - (-)^{a+b} \hat x_a \hat x_b\;.}

In section \ref{sec4} we construct the supersymmetric
extensions of the angular momentum and the Laplace-Runge-Lenz 
vector. Similarly as for the purely bosonic system, 
together they form a dynamical $SO(d+1)$ symmetry algebra.
This symmetry is exploited in the following section to 
obtain the discrete eigenvalues and their degeneracies.
In section \ref{sec6} we characterize all bound state wave 
functions. In the last section
we illustrate our general results by analyzing
in detail the supersymmetric hydrogen atoms in two, three and
four dimensions.

The appendices contain the necessary group theoretical tools needed in 
the main body of the paper.

\section{The Coulomb problem and its symmetries in $d$ Dimensions}
\label{sec2}
We generalize the Coulomb problem to arbitrary
dimensions by keeping the $1/r$-potential, although
this potential solves the Poisson equation in three dimensions
only. With this assumption the hydrogen atom in $d$ dimensions is
governed by the Schr\"odinger equation
\eqnl{H \psi = \left( -{\hbar^2\ov 2m} \triangle 
-{e^2\ov r} \right) \psi = E \psi\;.}{bos1}
It is convenient to measure distances in units
of the Compton wavelength $\lambdabar_c=\hbar/mc$. With respect to these dimensionless
coordinates equation \refs{bos1} takes the simpler form
\eqnl{
H\psi=E\psi\;,\qquad H=p^2-{\eta\ov r}\;,\qquad  p_a
=\frac 1 \im \, \pa_a\;, \qquad a=1,\dots,d\;,}{bos2}
where $\eta$ is twice the fine structure constant $\al$ and
the dimensionless energy $E$ is measured in units of $mc^2/2$.
The central force is attractive for positive $\eta$.

The hermitian generators $L_{ab}=x_a p_b -x_b p_a$ of the
rotation group in $d$ dimensions satisfy the 
familiar $so(d)$ commutation relations
\eqnl{
[ L_{ab}, L_{cd} ] = \im (\delta_{ac} L_{bd} 
+ \delta_{bd} L_{ac} - \delta_{ad} L_{bc} - \delta_{bc}
L_{ad} ) \;, }{bos5}
where indices run from $1$ to $d$.
It is not very difficult to guess the
generalization of the Laplace-Runge-Lenz vector \refs{einf2}
in $d$ dimensions \cite{sudarshan},
\eqnl{
C_a = L_{ab} p_b + p_b L_{ab} -{\eta x_a\ov r} \;.}{bos6}
Indeed, these operators commute with the Hamiltonian 
\refs{bos2}. They form a $SO(d)$-vector,
\[
[L_{ab},C_c]=\im (\delta_{ac}C_b-\delta_{bc}C_a)\;,\]
and the commutator of $C_a$ and $C_b$
is proportional to the angular momentum:
\eqnl{
[ C_a, C_b ] = -4\im L_{ab} H\;.}{bos7}
Now we may proceed as we did in three dimensions and define
on the negative energy $(E<0)$ subspace of the Hilbert 
space $\mathcal{H}=L_2(\mathbbm{R}^d)$ the hermitian operators
\eqnl{
K_a = {1\ov \sqrt{-4H}}\,C_a \;,\mtxt{with}[K_a,K_b] 
= \im L_{ab} \;.}{bos8}
The operators $\{L_{ab},K_a\}$
form a closed symmetry algebra of dimension $(d+1)d/2$. 
More explicitly, they can be combined to form generators 
$L_{AB}$ of the orthogonal group\footnote{For scattering
states $(E>0)$ a similar redefinition leads to 
generators of the Lorentz group $SO(d,1)$.
Here we are interested in bound states and will not 
further discuss this possibility.} $SO(d+1)$
\eqnl{
L_{AB} = \left(\begin{array}{c|c} L_{ab} & K_a \\ \hline 
-K_b & 0 \end{array}\right),}{bos9}
which implies that the $L_{AB}$ obey the commutation 
relations \refs{bos5} with indices running from
$1$ to $d+1$. We can verify a relation similar to
\refs{einf5} by solving
\eqnl{
 C_a C_a =-4K_aK_a H
=\eta^2 + \big(2 L_{ab} L_{ab} +(d-1)^2\big) H}{bos10}
for $H$. We obtain the
Hamiltonian
\eqnl{
H =p^2-{\eta \ov r}=
 - {\eta^2\ov (d-1)^2+4\, \mathcal{C}_{(2)}}}{bos12}
in terms of the second-order Casimir operator of the dynamical symmetry group $SO(d+1)$,
\[
\mathcal{C}_{(2)}= \frac 12 L_{AB} L_{AB} 
=\frac 12 L_{ab} L_{ab} + K_aK_a \;.\]

This is the generalization of \refs{einf5} we have
been looking for. It remains to characterize those
irreducible representations of $SO(d+1)$
which are realized in the Hilbert space $L_2(\R^d)$.
In three dimensions the allowed representations are
fixed by the condition $\bK\cdot\bL=0$
on the second Casimir operator of $SO(4)$. We expect $n\ms 1$
conditions on the $n$ Casimir operators 
of the dynamical symmetry group $SO(2n\ps 1)$ in $d=2n$ 
dimensions and $n$ conditions on 
the $n\ps 1$ Casimir operators of $SO(2n\ps 2)$ in $d=2n\ps 1$ 
dimensions. In the following we treat the even- and
odd-dimensional cases separately.

\textbf{Even-dimensional spaces:}
An irreducible representation is uniquely characterized
by its highest weight state. By definition, this state is annihilated by all
raising operators belonging to the simple roots.
To characterize these states one conveniently 
chooses complex coordinates $z_1,\dots,z_n$ in 
$\R^{2n}$ such that the raising operators of the 
dynamical symmetry group with generators $J_{AB}$ 
in \refs{bos9} have the simple form (cf. appendix \ref{append1})
\begin{eqnarray}
E_i&=&{1\ov \im}\big(z_i\pa_{i+1}-\bar z_{i+1}\bar\pa_i\big)\;,\quad i=1,\dots,n-1\;,
\label{bos13}\\
E_n&=&{1\ov\sqrt{2}}\big(C_{d-1}+\im C_d\big)=
-2z_n\Delta+(2r\pa_r+d-1)\bar\pa_n-{\eta z_n\ov r} \;.\label{bos14}
\end{eqnarray}
In the formula for $E_n$ we actually should have 
used the operators $K_{d-1}$ and $K_d$ in \refs{bos8},
since they appear as components of $J_{AB}$. But since
we are only interested in highest weight states 
which are annihilated by $E_n$ we may take the operators 
$C_{d-1}$ and $C_d$ instead.
Also note that all simple roots of $SO(d)$ are
positive roots of $SO(d\ps 1)$ so that all highest weight states 
of  $SO(d\ps 1)$ 
are automatically highest weight states of $SO(d)\subset SO(d\ps 1)$.
Now it is not difficult to see that a regular wave function
which is annihilated by all simple raising operators of $SO(d)$,
that is by the $n\ms 1$ operators in \refs{bos13} and by
the operator $z_{n-1}\bar\pa_n-z_n\bar\pa_{n-1}$ 
(see appendix \ref{append1a}), must have the form (cf. \refs{apen18})
\[ \Psi= f(r) z_1^\ell\;.\] 
It is a highest weight state of $SO(d+1)$, if in addition it is 
annihilated by $E_n$ in \refs{bos14}:
\eqnl{
E_n \Psi=-\left\{(d-1+2\ell){\ud\ov \ud r}\log f+\eta \right\}{z_n\ov r}\Psi
=0\;.}{bos16}
Hence the highest weight state reads
\eqnl{
\Psi=\ex^{-\gam_\ell r}\,z_1^\ell \;, \qquad \gam_\ell={\eta\ov
d-1+2 \ell}\;.}{bos17}
The constant $\eta$ must be positive for bound states to exist. 
$\Psi$ is of course an eigenfunction of all $n$
Cartan generators $H_i=z_i\pa_i-\bar z_i\bar\pa_i$ of $SO(2n+1)$
with eigenvalues $(\ell,0,\ldots,0)$. That shows that only
the symmetric multiplets of the dynamical 
symmetry group appear\footnote{This corresponds to the extension of
Fock's method to $d$ dimensions, cf. \cite{fock, alliluev, bander}.}. 
From \refs{apen14} we take the values of the second-order
Casimir operator for symmetric multiplets
\eqnl{
{\mathcal{C}_{(2)}}=
\ell ( \ell + d - 1) \;, \qquad  \ell = 0, 1, 2, \ldots}{bos18}
and their dimensions
\eqnl{
\hbox{dim}\, V_\ell=
{\ell + d\choose\ell}
-{\ell + d - 2 \choose \ell-2}\; .}{bos19}
The dim$V_\ell$ states of the symmetric representation
are obtained by acting repeatedly with the lowering operators
\[
E_{i<n}^\dagger=
{1\ov \im}\big(z_{i+1}\pa_i-\bar z_i\bar\pa_{i+1}\big)\mtxt{and}
E_n^\dagger=
-2\bar z_n\Delta+\pa_n(2r\pa_r-1)-{\eta \bar z_n\ov r} \]
on the state \refs{bos17}. This way one obtains
all bound $H$-eigenstates with the same energy.

\textbf{Odd-dimensional spaces:}
For $d=2n\ps 1$ the rank of the dynamical symmetry group $SO(d+1)$
exceeds the rank of the rotation group $SO(d)$ by one.
We combine the first $2n$ coordinates to $n$
complex ones and take coordinates $z_1,\dots,z_n,x_d$ in $\R^{d}$,
see appendix \ref{append1b}.
As Cartan generators we choose
\eqnl{
H_i=z_i\pa_i-\bar z_i\bar\pa_i \;, \quad i=1,\dots,n\mtxt{and}
H_{n+1}=C_d \;.}{bos21}
The raising operators are the $n-1$ operators $E_i$ in
\refs{bos13} supplemented by
\[
E_n={1\ov \im}\big(z_n\pa_{x_d}-x_d\bar\pa_{z_n}\big)
\mtxt{and}
E_{n+1}=\has(C_{d-2}+\im C_{d-1})\;.\]
The last raising operator $E_{n+1}$ coincides with $E_n$
in \refs{bos14}. A regular wave function is 
annihilated by the first $n$ raising operators only if 
it has the form $\Psi=f(r)z_1^\ell$.
The requirement that it is annihilated by the last
raising operator $E_{n+1}$ again leads to equation
\refs{bos16} and hence to the solution $\Psi$ in \refs{bos17}.
To determine the multiplets with this
highest weight state we need to calculate the highest weight
vector, that is the value of the Cartan generators on $\Psi$.
Clearly,
\eqnn{
H_1\Psi=\ell\Psi\mtxt{and} H_i\Psi=0\mtxt{for} i=2,\dots,n \;.}
The last Cartan generator in \refs{bos21} has the explicit
form
\[
H_{n+1}=-2x_d\triangle+(2r\pa_r+d-1)\pa_{x_d}-{\eta x_d\ov r}\]
and we find
\eqnn{
H_{n+1}\Psi=-\left\{(d-1+2\ell){\ud\ov \ud r}\log f+\eta\right\}{x_d\ov
r}\Psi =0 \;,}
after using $f=\ex^{-\gamma_{\ell}r}$. Hence, 
on any highest weight state the operators
$H_2,\dots,H_{n+1}$ vanish and again we find the
completely symmetric representations of the dynamical
symmetry group $SO(d+1)$. The eigenvalues of the second-order
Casimir operator and the dimension of the
representations are given by the same formulae
(\ref{bos18},\ref{bos19}) as for the hydrogen atom
in even dimensions.

Since $\eta$ is twice the fine structure
constant $\al=e^2/\hbar c$ and $E$ is measured in units of $mc^2/2$
the formula \refs{bos12} yields the well-known \cite{cornwell} bound state 
energies in three dimensions
\[
E_\ell= - {\al^2 \ov 1+\ell(\ell+2)}{mc^2\ov 2}
=-{me^4\ov 2\hbar^2}{1\ov n^2}\equiv E_n \;, \qquad n=1+\ell=1,2,\dots\;.\]
The degeneracy of $E_n$ is the dimension $n^2$ of the 
corresponding symmetric representation of $SO(4)$.
All $n^2$ states with the same energy $E_n$ are gotten 
by acting with the two lowering operators on 
the highest weight state
\eqnn{
\Psi(x)=\ex^{-\gam_n r}(x_1+ix_2)^\ell\;, \qquad
\gam_n={\al\ov n}\;.}
In $d$ dimensions the corresponding formulae read
\[E_\ell = -{me^4\ov 2\hbar^2}\Big(\ell+{d-1\ov 2}\Big)^{-2}\;,\qquad
\Psi(x)=\ex^{-\gam_\ell r}(x_1+\im x_2)^\ell\;, \qquad
\gam_\ell={\al\ov \ell+(d-1)/2}
\]
and $E_\ell$ has degeneracy dim $V_\ell$ in \refs{bos19}.

The appearance of the accidental degeneracy -- phrased in the language of representation theory --
corresponds to the following branching rule: the completely
symmetric representations of the dynamical symmetry group $SO(d+1)$
branches into those completely symmetric representations of the rotation group $SO(d)$ with 
equal or shorter Young diagrams,
{\small
\eqnl{
\quad \young(12\cdot\cdot \ell)\;\Bigg|_{SO(d+1)}\quad \longrightarrow
\quad\left\{\vphantom{\bigg|}\id\oplus\yng(1)\oplus \yng(2)
\oplus\dots\oplus \young(12\cdot\cdot\ell)\right\}\Bigg|_{SO(d)}\;,}{bos28}}
all of them possessing the same energy.
The energy, its degeneracy and the bound state wave
functions are uniquely fixed by the representation
of the dynamical symmetry group. Every completely symmetric
representation of $SO(d+1)$ appears once and only once
and corresponds to the multiplet with energy $E_\ell$. The angular 
momentum content of this multiplet is determined by the
branching rule \refs{bos28}.
In three dimensions this expresses just the well-known fact
that for each value of $E_{n=\ell+1}$ the orbital angular momentum
can vary from $0$ to $n-1=\ell$.

\section{$\cN=2$ Supersymmetric Quantum Mechanics}
\label{sec3}
We wish to further generalize our
results to the supersymmetric hydrogen atom
in $d$ dimensions. For that purpose we need
a supersymmetric extension of $d$-dimensional
Schr\"odinger operators and in particular of
the operator in \refs{bos2}. Such supersymmetric Hamiltonians 
can be written as
\eqnl{
H=\{Q,\Qd\}=H^\dagger\mtxt{with}Q^2=Q^{\dagger\,2}=0\;,}{sqm1}
where the supercharge $Q$ and its adjoint $\Qd$
anticommute with a self-adjoint idempotent operator $\Ga$.
The subspace on which $\Ga=1$ is called the bosonic sector and its
complement the fermionic sector. Hence, $Q$ transforms bosons
into fermions and vice versa. From \refs{sqm1} 
one sees at once that the supercharge
commutes with the supersymmetric Hamiltonian,
\[
[Q,H]=0 \;,\]
i.e. generates a supersymmetry of the system.
The simplest models exhibiting the structure 
\refs{sqm1} are $2\times 2$-dimensional matrix Schr\"{o}dinger operators 
in one dimension. Such models were first studied by 
Nicolai and Witten \cite{nicolai, witten1, witten2}.

Supersymmetric Hamiltonians in higher dimensions
have been introduced previously by several
authors \cite{cowi,andrianov}. Here we briefly present the 
construction used in this paper. 
We introduce a set of fermionic 
creation and annihilation operators, 
\eqnl{
\{\psia,\psibd\}=\delta_{ab} \;,\qquad\{\psia,\psib\}
=\{\psiad,\psibd\}=0\;,\qquad a,b=1,\dots,d}{sqm5}  
and the Fock space with vacuum $\vac$ which
is annihilated by all operators $\psia$. This space splits into sub-spaces, 
\[
\cC=\cC_0\oplus \cC_1\oplus \ldots \oplus \cC_d \;,\qquad 
\hbox{dim}\,\cC_p={d\choose p}\;,\quad
\hbox{dim}\,\cC=2^d\;,\]
labeled by their 'fermion number'
\[
N\vert_{\,\cC_p}=p\,\id\; ,\mtxt{where}
N=\sum_{a=1}^d \psiad\psia \; .\]
As basis in $\cC_p$ we may choose
\eqnl{
\vert a_1\dots a_p\ra=\psid_{a_1} \ldots \psid_{a_p}\vac\;,\qquad
a_1<a_2< \ldots <a_p \;.}{states}
Along with $\cC$ the Hilbert space of all square integrable 
wave functions decomposes as 
\eqnl{
\cH=\cH_0\oplus\cH_1\oplus \ldots \oplus\cH_d \;, \mtxt{where}
N\big|_{{ \cal H}_p}=p \,\unitop \;.}{sqm8}
An arbitrary wave function in $\cH_p$ has the
expansion
\eqnl{
\Psi=f_{a_1 \ldots a_p}(x)\vert a_1\dots a_p\ra \;,\qquad
f_{a_1\ldots a_p} \mtxt{totally antisymmetric.}}{sqm8a}
An explicit realization of the creation and annihilation
operators can be given in terms of the hermitian 
$\gam$-matrices in $2d$ Euclidean dimensions:
$\psia=\ha (\ga_a-\im\ga_{d+a})$.

The supercharge and its adjoint\footnote{The hermitian linear 
combinations $Q_1 = Q + \Qd$ and $Q_2 = \im (Q - \Qd)$
satisfy the standard $\cN=2$ supersymmetry algebra 
$\{ Q_i, Q_j \} = 2 H \delta_{ij}$.}
are defined via
\eqngrl{
Q & = & \ex^{-\chi}Q_0 \ex^\chi
=\im \sum_a \psi_a (\pa_a + \pa_a \chi)\;,\mtxt{with}
Q_0=\im\psi_a\pa_a\;,}
{\Qd& = & \ex^{\chi}\Qd_0 \ex^{-\chi}
=\im \sum_a \psi_a^\dagger (\pa_a -\pa_a \chi)\;,\mtxt{with}
\Qd_0=\im\psid_a\pa_a \;.
}{sqm9}
At this point the real superpotential $\chi(x_1,\dots,x_d)$
remains unspecified. From (\ref{sqm5}) it follows
at once that the free supercharge $Q_0$ is nilpotent and
since $Q$ and $Q_0$ are related by a similarity 
transformation the same holds true for $Q$.
The supercharge $Q$ only contains annihilation operators
and hence decreases the fermion number by one. Its adjoint
$\Qd$ increases it by one,
\[
[N,Q]=-Q\mtxt{and}
[N,\Qd]=\Qd\;.\]
The  supersymmetric Hamiltonian defined in \refs{sqm1} is
a $2^d\times 2^d$-dimensional matrix Schr\"odinger operator and
takes the following form (cf. also 
\cite{cowi,andrianov})
\eqnl{
H= \big\{-\triangle + (\nabla \chi,\nabla\chi) 
+ \triangle \chi\big\}\id_{2^d}
- 2 \sum_{a,b=1}^d\psi^\dagger_a\,\chi_{ab}\, \psi_b \;,\quad
\chi_{ab}={\pa^2\chi\ov\pa x_a\pa x_b}\;.
}{sqm10}
We use brackets to indicate contraction of indices as 
$(A,B)=\sum_a A^a B^a$. 
Contrary to the supercharge and its adjoint the 
supersymmetric Hamiltonian $H$
commutes with the number operator $N$ and 
hence leaves each subspace $\cH_p$ in the 
decomposition \refs{sqm8} invariant,
\eqnn{
H:\cH_p\longrightarrow\cH_p \;.}
On the subspace $\cH_p$ the supersymmetric Hamiltonian
is still a matrix Schr\"odinger operator,
\[
H\big\vert_{\cH_p}=-\triangle\id+V^{(p)}\;,\qquad
\tr \id={d\choose p}\;.\]
Only in the extreme sectors $\cH_0$ and $\cH_p$ 
do we get ordinary  Schr\"odinger operators acting on one-component 
wave functions. With
\eqnn{
\psid_a\psib\vac=0\mtxt{and} \psid_a\psib\;\vert 12\dots d\ra
=\delta_{ab}\vert 12\dots d\ra}
the corresponding potentials take the form
\[
V^{(0)} =  (\nabla \chi, \nabla \chi) + \triangle \chi
\mtxt{and}
V^{(d)} =  (\nabla \chi, \nabla \chi) - \triangle \chi\;.\]
More generally, for 
an arbitrary state $\Psi=f_{a_1\dots a_p}\vert a_1\dots a_p\ra\in \cH_p$
the Hamiltonian acts as follows:
\eqnl{
\la a_1\dots a_p\vert H\Psi\ra=(-\triangle+V^{(0)})f_{a_1\dots a_p}
+2\sum_{b,i=1}^p (-)^i\chi_{a_i b}\,f_{ba_1\ldots\check{a}_i\ldots a_p}\;.
}{sqm12}
The nilpotent supercharges give rise to the following 
Hodge-type decomposition of the Hilbert space,
\eqnl{
\cH=Q\cH \oplus \Qd\cH \oplus \hbox{Ker}\,H\;,}{sqm13}
where the finite dimensional subspace Ker$\,H$
is spanned by the zero-modes of $H$. Indeed,
on the orthogonal complement of Ker$\,H$ we may invert $H$
and write
\eqnn{
\cH_0^\perp=(Q\Qd+\Qd Q)H^{-1}\cH_0^\perp
=Q\Big({\Qd\ov H}\cH_0^\perp\Big)
+\Qd \Big({Q\ov H}\cH_0^\perp\Big)\;,}
which proves \refs{sqm13}.
The supercharge  $Q$ maps every energy-eigenstate
in $\Qd\cH\cap\cH_p$
with positive energy into an eigenstate
in $Q\cH\cap\cH_{p-1}$ with the same energy. 
Its adjoint maps eigenstates in $Q\cH\cap \cH_p$ into 
those in $\Qd\cH\cap \cH_{p+1}$ with the same
energy. With the exception of the zero-energy 
states there is an \emph{exact pairing} between the eigenstates and 
energies in the bosonic and in the fermionic sector as
depicted below.

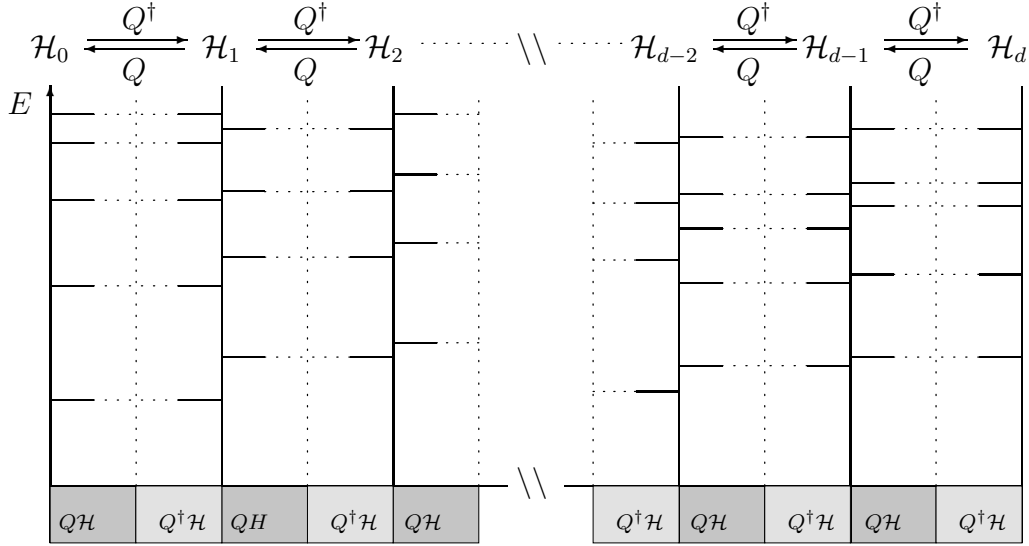
\setlength{\unitlength}{0,038mm}
\begin{center}
\begin{picture}(4000,1900)(0,0)

\multiput(200,0)(600,0){3}{\line(0,1){1600}}
\put(200,1500){\vector(0,1){100}}
\put(50,1500){$E$}
\put(130,1700){$\cH_0$}
\put(330,1760){\vector(1,0){350}}
\put(670,1730){\vector(-1,0){350}}
\put(450,1800){$\Qd$}
\put(450,1620){$Q$}
\put(730,1700){$\cH_1$}

\put(930,1760){\vector(1,0){350}}
\put(1270,1730){\vector(-1,0){350}}
\put(1050,1800){$\Qd$}
\put(1050,1620){$Q$}
\put(1300,1700){$\cH_2$}
\texture{44000000 aaaaaa aa000000 8a888a 88000000 aaaaaa aa000000 888888 
         88000000 aaaaaa aa000000 8a8a8a 8a000000 aaaaaa aa000000 888888 
         88000000 aaaaaa aa000000 8a888a 88000000 aaaaaa aa000000 888888 
         88000000 aaaaaa aa000000 8a8a8a 8a000000 aaaaaa aa000000 888888 }

\shade\path(200,0)(500,0)(500,200)(200,200)(200,0)
\shade\path(800,0)(1100,0)(1100,200)(800,200)(800,0)
\shade\path(1400,0)(1700,0)(1700,200)(1400,200)(1400,0)
\shade\path(2400,0)(2700,0)(2700,200)(2400,200)(2400,0)
\shade\path(3000,0)(3300,0)(3300,200)(3000,200)(3000,0)

\multiput(2400,0)(600,0){3}{\line(0,1){1600}}
\multiput(500,0)(600,0){3}{\dottedline{40}(0,200)(0,1550)}
\multiput(2100,0)(600,0){3}{\dottedline{40}(0,200)(0,1550)}
\put(2230,1700){$\cH_{d-2}$}
\put(2830,1700){$\cH_{d-1}$}
\put(3480,1700){$\cH_d$}

\put(2520,1760){\vector(1,0){290}}
\put(2810,1730){\vector(-1,0){290}}
\put(2600,1800){$\Qd$}
\put(2600,1620){$Q$}

\put(3120,1760){\vector(1,0){290}}
\put(3410,1730){\vector(-1,0){290}}
\put(3200,1800){$\Qd$}
\put(3200,1620){$Q$}
\put(200,200){\line(1,0){1600}}
\put(2000,200){\line(1,0){1600}}
\put(1830,180){$\backslash\backslash$}
\dottedline{40}(1500,1750)(1800,1750)
\dottedline{40}(1980,1750)(2200,1750)
\put(1830,1700){$\backslash\backslash$}

\texture{8101010 10000000 444444 44000000 11101 11000000 444444 44000000 
	101010 10000000 444444 44000000 10101 1000000 444444 44000000 
	101010 10000000 444444 44000000 11101 11000000 444444 44000000 
	101010 10000000 444444 44000000 10101 1000000 444444 44000000 }
\shade\path(500,0)(800,0)(800,200)(500,200)(500,0)
\shade\path(1100,0)(1400,0)(1400,200)(1100,200)(1100,0)
\shade\path(2100,0)(2400,0)(2400,200)(2100,200)(2100,0)
\shade\path(2700,0)(3000,0)(3000,200)(2700,200)(2700,0)
\shade\path(3300,0)(3600,0)(3600,200)(3300,200)(3300,0)

\multiput(200,500)(450,0){2}{\line(10,0){150}}
\put(350,500){\dottedline{40}(0,0)(300,0)}
\multiput(200,900)(450,0){2}{\line(10,0){150}}
\put(350,900){\dottedline{40}(0,0)(300,0)}
\multiput(200,1200)(450,0){2}{\line(10,0){150}}
\put(350,1200){\dottedline{40}(0,0)(300,0)}
\multiput(200,1400)(450,0){2}{\line(10,0){150}}
\put(350,1400){\dottedline{40}(0,0)(300,0)}
\multiput(200,1500)(450,0){2}{\line(10,0){150}}
\put(350,1500){\dottedline{40}(0,0)(300,0)}

\multiput(800,650)(450,0){2}{\line(10,0){150}}
\put(950,650){\dottedline{40}(0,0)(300,0)}
\multiput(800,1000)(450,0){2}{\line(10,0){150}}
\put(950,1000){\dottedline{40}(0,0)(300,0)}
\multiput(800,1230)(450,0){2}{\line(10,0){150}}
\put(950,1230){\dottedline{40}(0,0)(300,0)}
\multiput(800,1450)(450,0){2}{\line(10,0){150}}
\put(950,1450){\dottedline{40}(0,0)(300,0)}

\put(1400,700){\line(1,0){150}}
\put(1550,700){\dottedline{40}(0,0)(150,0)}
\put(1400,1050){\line(1,0){150}}
\put(1550,1050){\dottedline{40}(0,0)(150,0)}
\put(1400,1290){\line(1,0){150}}
\put(1550,1290){\dottedline{40}(0,0)(150,0)}
\put(1400,1500){\line(1,0){150}}
\put(1550,1500){\dottedline{40}(0,0)(150,0)}

\put(2250,530){\line(1,0){150}}
\put(2100,530){\dottedline{40}(0,0)(150,0)}
\put(2250,990){\line(1,0){150}}
\put(2100,990){\dottedline{40}(0,0)(150,0)}
\put(2250,1190){\line(1,0){150}}
\put(2100,1190){\dottedline{40}(0,0)(150,0)}
\put(2250,1400){\line(1,0){150}}
\put(2100,1400){\dottedline{40}(0,0)(150,0)}

\multiput(2400,620)(450,0){2}{\line(10,0){150}}
\put(2550,620){\dottedline{40}(0,0)(300,0)}
\multiput(2400,910)(450,0){2}{\line(10,0){150}}
\put(2550,910){\dottedline{40}(0,0)(300,0)}
\multiput(2400,1100)(450,0){2}{\line(10,0){150}}
\put(2550,1100){\dottedline{40}(0,0)(300,0)}
\multiput(2400,1220)(450,0){2}{\line(10,0){150}}
\put(2550,1220){\dottedline{40}(0,0)(300,0)}
\multiput(2400,1420)(450,0){2}{\line(10,0){150}}
\put(2550,1420){\dottedline{40}(0,0)(300,0)}

\multiput(3000,650)(450,0){2}{\line(10,0){150}}
\put(3150,650){\dottedline{40}(0,0)(300,0)}
\multiput(3000,940)(450,0){2}{\line(10,0){150}}
\put(3150,940){\dottedline{40}(0,0)(300,0)}
\multiput(3000,1180)(450,0){2}{\line(10,0){150}}
\put(3150,1180){\dottedline{40}(0,0)(300,0)}
\multiput(3000,1260)(450,0){2}{\line(10,0){150}}
\put(3150,1260){\dottedline{40}(0,0)(300,0)}
\multiput(3000,1450)(450,0){2}{\line(10,0){150}}
\put(3150,1450){\dottedline{40}(0,0)(300,0)}

\put(230,50){\scriptsize$Q\cH$}
\put(580,50){\scriptsize$\Qd\cH$}
\put(830,50){\scriptsize$QH$}
\put(1180,50){\scriptsize$\Qd\cH$}
\put(1440,50){\scriptsize$Q\cH$}
\put(2180,50){\scriptsize$\Qd\cH$}
\put(2440,50){\scriptsize$Q\cH$}
\put(2780,50){\scriptsize$\Qd\cH$}
\put(3050,50){\scriptsize$Q\cH$}
\put(3380,50){\scriptsize$\Qd\cH$}
\end{picture}
\end{center}

The supersymmetric systems with supercharges
\refs{sqm9} admit a generalized Poincar\'{e} duality
relating $\cH_{d-p}$ with $\cH_p$. This is seen
as follows: instead of the vacuum $\vac\in\cH_0$, which
is annihilated by all $\psi_a$, we take $\|0\ra
=\psid_1\cdots\psid_d\vac\in \cH_d$,
which is annihilated by all $\psid_a$.
As basis in $\cC_{d-p}$ we choose
\eqnn{
\| a_1\dots a_p\ra=\psi_{a_1}\cdots \psi_{a_p}\| 0\ra \;,} 
such that an arbitrary wave function in
$\cH_{d-p}$ has the expansion
\eqnl{
\Psi=f_{a_1\dots a_p}\|a_1\dots a_p\ra,\qquad f_{a_1\dots a_p}
\mtxt{totally antisymmetric.}}{sqm14}
The supersymmetric Hamiltonian acts on such a
state as follows,
\[
\la a_1 \ldots a_p \|H\Psi\ra=
(-\triangle+V^{(d)})f_{a_1\dots a_p}
-2\sum_{b,i=1}^p (-)^i\chi_{a_i b}\,f_{ba_1\ldots\check{a}_i\ldots a_p}\;.\]
A comparison with \refs{sqm12} yields the important duality
relation
\eqnl{
H^{\chi}\Big\vert_{\cH_p}=H^{-\chi}\Big\vert_{\cH_{d-p}}\;,}{sqm16}
which states, that the Hamiltonians in $\cH_p$ and $\cH_{d-p}$ are equivalent,
up to a sign-change of the superpotential.
Similarly on finds, that the action of $Q$ on
the state \refs{states} in $\cH_p$ is identical to
the action of $\Qd$ on the state \refs{sqm14}
in $\cH_{d-p}$, up to a sign-change of $\chi$.

\section{The supersymmetric Hydrogen Atom and its Symmetries}
\label{sec4}
First we consider $d$-dimensional supersymmetric 
systems with spherically symmetric superpotentials
and derive the conserved angular momentum $J_{ab}$.
The total angular momentum $J_{ab}$ is the sum of 
two terms: the orbital part $L_{ab}$ and the internal
part $S_{ab}$ which transforms the components of
a wave function.
For a superpotential $\chi(r)$ the supercharges 
simplify to
\eqnl{
Q=\im\psi_a\big(\pa_a+x_a f\big)\mtxt{and}
\Qd=\im\psid_a\big(\p_a-x_a f\big) \;, \mtxt{where} 
f = \frac {\chi^\prime}{r} \;.  } {lrl1}
They should be scalars with respect to the
rotation group $SO(d)$. However, since $\pa_a$ 
is a vector and $\psia$ a scalar with respect to $L_{ab}$, we need
to supplement orbital rotations by internal
ones such that $\psi_a$ becomes a vector as well.
One easily verifies that
\eqnl{
[S_{ab},\psi_c] = \im ( \delta_{ac} \psi_b - \delta_{bc} \psi_a )\;, 
\mtxt{where}
S_{ab}={1\ov \im}(\psid_a\psi_b-\psid_b\psi_a) \; ,}{lrl2}
and that the hermitian $S_{ab}$ satisfy the 
same commutation relations \refs{bos5} as the orbital angular
momentum. It follows at once from \refs{lrl2}
that the supercharge \refs{lrl1} commutes 
with the total angular momenta
\eqnl{
J_{ab}=L_{ab}+S_{ab}\; ,}{lrl3}
since $Q$ only contains scalar products of vectors operators.

Next we prove that there exists a supersymmetric 
generalization of the Laplace-Runge-Lenz 
vector in $d$ dimensions if the potential is $\sim 1/r$. 
For such a potential the supercharges \refs{lrl1} do
commute with the supersymmetric generalization of \refs{bos6},
\eqnl{
C_a=J_{ab}p_b+p_b J_{ab}+x_a f(r)A \;}{lrl4}
with a suitable operator $A$. One can show
that $f$ must be the function in \refs{lrl1}
for $C_a$ to commute with $Q$.
This function, along with $A$ are
fixed as follows:

First, for $C_a$ to be a vector, the operator $A$ must be a scalar 
under rotations induced by the $J_{ab}$.
Second, in the zero particle sector $\cH_0$ the vector
$C_a$ must coincide with $C_a$ in \refs{bos6}.
Third, $A$ should commute with the particle number operator
since the $J_{ab}$ do commute.

A general ansatz for $A$ subject to these
three conditions reads
\eqnl{
A=\al\id-\beta N-\gam S^\dagger S\;,\qquad
S=\hat x_a\psi_a\;,\quad \hat x_a={x_a\ov r}\;,}{lrl5}
with constants $\al,\beta$ and $\gam$ which ought to be determined. 
Clearly, $C_a$ is a vector operator, such that
\[
[J_{ab},C_c]=\im \big(\delta_{ac}C_b-\delta_{bc}C_a\big)\]
holds true. 
Let us now calculate the commutators between the 
$C_a$ and the supercharge. They should vanish for 
a suitable chosen function $f$ in
\refs{lrl4}. We obtain
\eqngrl{[C_a,Q]&=&
2\big\{f \psi_b+f'S x_b \big\} J_{ab}
+ \beta f x_a Q_0
+\im fx_a\big\{(\beta+\gam)rf+\gam \pa_r\big\}S}
{&+&\im\big\{f\psi_a+ f'S x_a\big\}\left(1-d-A\right)
+\im \gam\hat x_a f\,(d\ms N\ms 1)S \; .}{lrl7}
The terms containing derivatives are
\[
fx_a(\beta-2)Q_0+2\im r(f+rf')S\p_a
+\im (\gam f-2rf')S x_a\pa_r \; .\]
They cancel if
\eqnl{
f= - {\lam\ov r}\mtxt{or}
\chi= - \lam r\mtxt{and} \beta=-\gam=2}{lrl8}
hold true. With this choice all but the terms
proportional to $S_{ab}$ in the first line on the right 
in \refs{lrl7} cancel, and we remain with
\[
[C_a,Q]={\im\lambda\ov r}\, (\al -d +1)
\big(\psi_a-\hat x_a S\big)\]
which vanishes for $\alpha=d-1$.
Hence, the supersymmetric extension of the
conserved Laplace-Runge-Lenz vector reads
\eqnl{
C_a = J_{ab}p_b+p_b J_{ab}-\lam\hat x_a A\mtxt{with}
A =(d-1)\id -2N+2S^\dagger S\;,
}{lrl10}
and this is the main result of this section.
The choice $\chi = - \lambda r$ for the superpotential 
in \refs{sqm9} and \refs{sqm10} leads to 
the following supersymmetric extension of the Coulomb Hamiltonian
\eqnl{
H= - \triangle + \lambda^2 - \frac {\lambda A}{r} \; .}{lrl11}
Restricted to the zero particle sector this is the 
Hamiltonian of the hydrogen 
atom\footnote{We have to identify $\eta=\lambda(d-1)$. 
The additional shift $\lambda^2$ makes the lowest eigenvalue 
of this operator to be equal to zero.} and restricted to the $d$ particle 
sector it corresponds to the electron-antiproton scattering system.
The corresponding supercharge and its adjoint take the simple form
\[
Q=Q_0-\im\lam S\mtxt{and}
\Qd=\Qd_0+\im\lam S^\dagger \;,\]
where the free supercharge has been defined in \refs{sqm9}
and the operator $S=\hat x_a\psi_a$ has already
been used in \refs{lrl5}.

\section{Algebraic Determination of the Spectrum}
\label{sec5}

We proceed as we did in the purely bosonic case and
calculate the commutator of two $C_a$:
\[
[C_a,C_b]=-4\im J_{ab}\left( -\triangle - {\lam\ov r}A \right)
\stackrel{\refs{lrl11}}{=}-4 \im J_{ab}\left( H-\lam^2 \right) \; .\]
Up to the shift in $H$ and the replacement $L_{ab} \to J_{ab}$ 
this is the same relation as (\ref{bos7}). The
total angular momenta $J_{ab}$ and the vector operator
\eqnl{
K_a = \frac {C_a}{\sqrt{4(\lambda^2 - H)}}}{alg2}
form a $SO(d+1)$ algebra on the subspace of bound states $(E<\lambda^2)$. 
Finally we should calculate $C_aC_a$. If we can express
this scalar operator in terms of $H$ and operators that 
commute with $H$, similarly as we did in \refs{bos10},
then we may solve for $H$. However, one soon realizes that
this is impossible by only using the operators $\id, N, J_{ab}J_{ab}$
and $H$. However, we can express $C_aC_a$ in
terms of $\id,N,J_{ab}J_{ab}$ and the two operators
$Q\Qd$ and $\Qd Q$ as follows:
\eqngrl{
C_a C_a=4(\lam^2-H)K_aK_a=
-2\lam^2 J_{ab}J_{ab}
&+&\left( 2 J_{ab} J_{ab} + (d-2N-1)^2 \right) Q\Qd}
{&+& \left( 2J_{ab} J_{ab}
+(d - 2N +1)^2 \right) \Qd Q \;,}
{alg3}
and this relation is sufficient to obtain the
spectrum of the supersymmetric hydrogen atom.

Each of the three subspaces in the decomposition
\refs{sqm13} is left invariant by the Hamiltonian
and we may consider $H$ on each subspace separately.
Since $Q^2=0$ we find $H\vert_{Q\cH}=Q\Qd$ 
and $H\vert_{\Qd\cH}=\Qd Q$ and can solve 
\refs{alg3} for $H$ in each of these subspaces
\begin{eqnarray}
H\big\vert_{Q\cH}&=&
Q\Qd=\lam^2-{(d-2N-1)^2\lam^2\ov (d-2N-1)^2+4\cC_{(2)}}\;,\label{alg5a}\\
H\big\vert_{\Qd\cH}&=&
\Qd Q=\lam^2-{(d-2N+1)^2\lam^2\ov (d-2N+1)^2+4\cC_{(2)}}\;,\label{alg5b}
\end{eqnarray}
where $\cC_{(2)}$ is the second-order Casimir
of the dynamical symmetry group $SO(d\ps 1)$,
\[
\cC_{(2)} = {1\ov 2}J_{AB} J_{AB} 
={1\ov 2} J_{ab} J_{ab} + K_aK_a \; .\]
All zero modes of $H$ are annihilated by both $Q$ and
$\Qd$, and according to \refs{alg3} 
the second-order Casimir must vanish on these modes,
such that
\[
\cC_{(2)}\big\vert_{\hbox{Ker}\,H}=0\;.\]
We conclude that every normalizable zero mode $\Psi$ of $H$ must transform
trivially under the dynamical symmetry group, $J_{AB}\Psi=0$.

To obtain the bound state energies we need
to determine those irreducible representations
of the dynamical symmetry group
which are realized in $\cH$ and the corresponding
values of the second-order Casimir operator. The 
degeneracy of an energy level is then equal to the dimension 
of the corresponding representation.

We use the abbreviation $\cD{\ell}{\wp}$ to denote 
multiplets of the orthogonal groups
corresponding to Young tableaux of the form
\eqnn{
\young(1\cdot\cdot\ell,\cdot,\cdot,\wp)}
since in the following only those representations will appear.
Let us assume, that each component function
$f_{a_1\dots a_p}(x)$, entering the state $\Psi\in\cH_p$ in \refs{sqm8a},
is a harmonic polynomial of degree $\ell$, that
is
\eqnn{f_{a_1\dots a_p}(x)=
\sum_{b_1,b_2,\dots ,b_\ell}f_{a_1\dots a_p b_1\dots b_\ell}
x_{b_1}x_{b_2}
\cdots x_{b_\ell}}
with $f$ symmetric in the $b-$indices and of zero trace in
each pair of them. 
Since $f_{a_1\dots a_p}$
is completely antisymmetric in the $a-$indices, these
objects transform according to the
completely antisymmetric representation
\[
\cD{1}{\wp}\sim \young(1,\cdot,\cdot,\wp)\]
of the rotation group $SO(d)$ generated by the
$S_{ab}$. 
On the other hand, each homogeneous polynomial
$f_{a_1\dots a_p}(x)$ transforms according
to the completely symmetric representations
\[
\cD{\ell}{1}\sim\young(1\cdot\cdot \ell)\]
of the rotation group generated by the $L_{ab}$.
It follows, that the wave function $\Psi\in\cH_p$
transforms according to the tensor-product representations
\eqnl{
\cD{1}p
\otimes \cD{\ell}1
= \cD{\ell}{p-1} \oplus \cD{\ell-1}p \oplus \cD{\ell+1}p
\oplus \cD{\ell}{p+1} \;.}{alg10}
Recall, that $p$ is the fermion number and $\ell$ the order
of the homogeneous polynomials.
For $\psi\in\cH_0$ or $\psi\in\cH_d$ the first
factor on the left hand side in
\refs{alg10} becomes the trivial representation and we
only obtain the fully symmetric representations $\cD\ell 1$
on the right hand side, in agreement with our earlier results in the purely bosonic
case. In the sectors $\cH_1$ and $\cH_{d-1}$
the first representation on the right hand side
of \refs{alg10} is absent. For linear functions with $\ell=1$ the
second representation on the right is missing.
Finally, when using the results \refs{alg10}, one 
should keep in mind that the representations 
with $\wp$ and $d\ms\wp$ of $SO(d)$ are equivalent,
$\cD{\ell}{\wp} \sim \cD{\ell}{d-\wp}\;,$
and that for even dimensions the representations
$\cD\ell{d/2}$ are reducible and contain
one selfdual and one anti-selfdual multiplet.
All representations of the rotation group
$SO(d)$  appearing in 
the tensor product \refs{alg10} should
group together into multiplets
of the dynamical symmetry group
$SO(d\ps 1)$. To continue we need the following
rules for the branching of $SO(d\ps 1)$- into
$SO(d)$-representations\footnote{They can be obtained from \cite{mckay} or by using the program LiE.},
\begin{eqnarray}
\label{bra1}
\cD{\ell}{\wp} \;\Big|_{SO(d+1)} \!\!\longrightarrow 
\left\{\cD{\ell}{\wp} \oplus \cD{\ell-1}{\wp} \oplus \ldots
\oplus \cD{1}{\wp}\;\;\; \oplus \;\;\;\cD{\ell}{\wp-1} 
\oplus \cD{\ell-1}{\wp-1} \oplus \ldots
\oplus \cD{1}{\wp-1} \right\} \Big|_{SO(d)}\;.
\end{eqnarray}
Now it is not difficult to see, that in the sector 
$\cH_p$ all polynomials up to order $\ell$ appear in the
branching of only two $SO(d+1)$-multiplets
up to two extreme representations,
\begin{eqnarray*}
\cD{\ell}{\wp} \oplus \cD{\ell}{\wp+1} \Big|_{SO(d+1)}
\!\!\longrightarrow \left\{ \cD{1}{\wp} \otimes
\left( \unitop \oplus \cD{1}{1} \oplus \ldots \oplus \cD{\ell}{1} \right)
 - \cD{\ell+1}{\wp} + \cD{\ell}{\wp-1} \right\} \Big|_{SO(d)}
\;.
\end{eqnarray*}
Of course, for $\wp=1$ the last representation of the
rotation group is absent. There is one notable
exception to these branching rules for even $d$: 
in the middle sector $\cH_{n=d/2}$ the correct branching rule reads
\[
\cD{\ell}{n} \oplus \cD{\ell}{n} \Big|_{SO(d+1)} 
\!\!\longrightarrow \left\{ \cD{1}{n} \otimes
\left( \unitop \oplus \cD{1}{1} \oplus \ldots \oplus \cD{\ell}{1} \right) 
- \cD{\ell+1}{n} -\cD{\ell}{n} - \cD{\ell}{n-1} \right\} \Big|_{SO(d)}
\;.\]
We summarize our results:
In \emph{odd dimensions} $d=2n\ps 1$ the following representations
of $SO(2n\ps 2)$ containing bound states arise in the various
sectors of $\cH$ for $\ell\geq 1$:
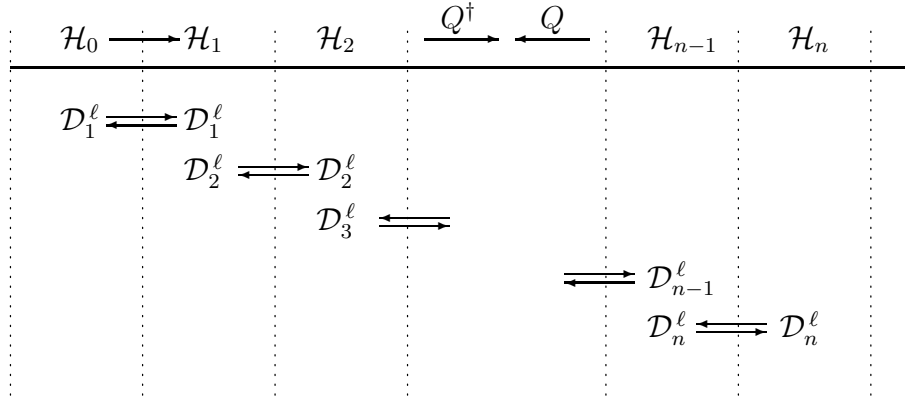
\setlength{\unitlength}{0,022mm}\label{sectors1}
\begin{center}
\begin{picture}(5600,2300)(0,700)
\multiput(0,200)(800,0){4}{\dottedline{60}(0,400)(0,2600)}
\multiput(3600,200)(800,0){3}{\dottedline{60}(0,400)(0,2600)}
\put(0,2580){\line(1,0){5400}}

\put(300,2700){$\cH_0$}
\put(600,2750){\vector(1,0){420}}
\put(1050,2700){$\cH_1$}
\put(1850,2700){$\cH_2$}
\put(2500,2750){\vector(1,0){450}}
\put(2600,2820){$\Qd$}
\put(3500,2750){\vector(-1,0){450}}
\put(3200,2820){$Q$}
\put(4700,2700){$\cH_n$}
\put(3850,2700){$\cH_{n-1}$}
\put(300,2200){$\cD\ell 1$}
\put(580,2280){\vector(1,0){420}}
\put(1000,2230){\vector(-1,0){420}}
\put(1050,2200){$\cD\ell 1$}

\put(1050,1900){$\cD\ell 2$}
\put(1800,1930){\vector(-1,0){420}}
\put(1380,1980){\vector(1,0){420}}
\put(1850,1900){$\cD\ell 2$}

\put(1850,1600){$\cD\ell 3$}
\put(2230,1620){\vector(1,0){420}}
\put(2650,1670){\vector(-1,0){420}}

\put(3350,1330){\vector(1,0){420}}
\put(3770,1280){\vector(-1,0){420}}

\put(3850,1250){$\cD\ell {n-1}$}
\put(3850,950){$\cD\ell n$}
\put(4570,1030){\vector(-1,0){420}}
\put(4150,980){\vector(1,0){420}}

\put(4650,950){$\cD\ell n$}
\end{picture}
\end{center}
In all sectors but $\cH_0$ we have $\ell\in \N$. 
In the subspaces $\cH_0$ we have $\ell\in\N_0$
and $\ell=0$ corresponds to the trivial representation.
The sectors $\cH_{p>n}$ support no bound states
(see below)  and therefore it suffices to consider the 
sectors with $N\leq n$. 

In \emph{even dimensions} $d=2n$
the following representations of the dynamical symmetry
group $SO(2n\ps 1)$ arise for $\ell\geq 1$:

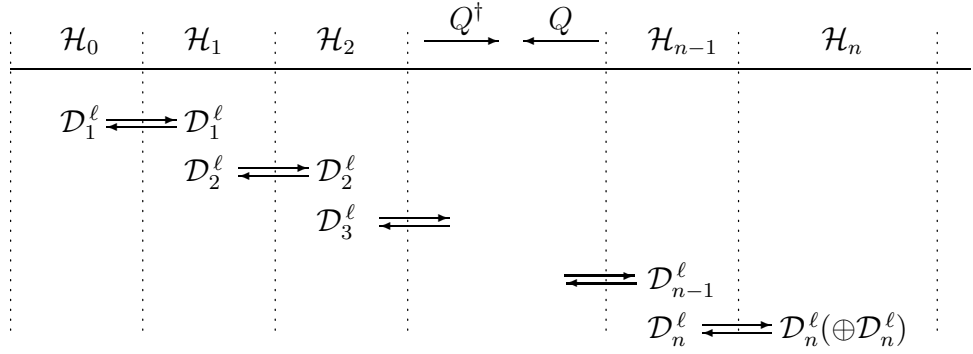
\setlength{\unitlength}{0,022mm}\label{sectors2}
\begin{center}
\begin{picture}(5900,2200)(0,700)
\multiput(0,200)(800,0){4}{\dottedline{60}(0,800)(0,2600)}
\multiput(3600,200)(800,0){2}{\dottedline{60}(0,800)(0,2600)}
\dottedline{60}(5600,1000)(5600,2800)
\put(0,2580){\line(1,0){5800}}

\put(300,2700){$\cH_0$}
\put(1050,2700){$\cH_1$}
\put(1850,2700){$\cH_2$}
\put(2500,2750){\vector(1,0){450}}
\put(2650,2820){$\Qd$}
\put(3550,2750){\vector(-1,0){450}}
\put(3250,2820){$Q$}
\put(3850,2700){$\cH_{n-1}$}
\put(4900,2700){$\cH_n$}

\put(300,2200){$\cD\ell 1$}
\put(580,2270){\vector(1,0){420}}
\put(1000,2230){\vector(-1,0){420}}
\put(1050,2200){$\cD\ell 1$}

\put(1050,1900){$\cD\ell 2$}
\put(1800,1930){\vector(-1,0){420}}
\put(1380,1980){\vector(1,0){420}}
\put(1850,1900){$\cD\ell 2$}

\put(1850,1600){$\cD\ell 3$}
\put(2230,1680){\vector(1,0){420}}
\put(2650,1630){\vector(-1,0){420}}
\put(3350,1330){\vector(1,0){420}}
\put(3780,1280){\vector(-1,0){420}}
\put(3850,1250){$\cD\ell {n-1}$}
\put(3850,950){$\cD\ell n$}
\put(4600,980){\vector(-1,0){420}}
\put(4180,1030){\vector(1,0){420}}
\put(4650,950){$\cD\ell n(\oplus \cD\ell n)$}

\end{picture}
\end{center}
The values of the quadratic Casimir operator
of $SO(d+1)$ entering the formulae (\ref{alg5a},\ref{alg5b})
for the energies are
\eqnl{\cC_{(2)} \left( \cD{\ell}{\wp} \right) = d (\ell+\wp-1)+\ell(\ell-1)-\wp(\wp-1) \;.}{alg14}
The dimensions of the representations $\cD\ell \wp$
are given in the appendix. In odd dimensions one
uses the formula 
\refs{apen11} with $2n$ replaced by $2n=d\ps 1$ 
and in even dimensions the formula \refs{apen25} 
with $d\ps 1=2n\ps 1$. In additions one must
set 
\eqnn{
(\ell_1,\ell_2,\dots,\ell_n)=(\overbrace{\ell,1,\dots,1}^p,0\dots,0)}
in these formulae.

Now we are ready to determine all eigenvalues
of the supersymmetric Hamiltonian \refs{lrl11}
with the help of the results (\ref{alg5a},\ref{alg5b}) and \refs{alg14} 
as follows:

In $\cH_0$ only the symmetric representations $\cD\ell 1$
of $SO(d\ps 1)$ are realized. Since in addition $Q\vert_{\cH_0}=0$
we obtain the following eigenvalues for $H$ in
\refs{alg5a}
\[
\cH_0: \qquad E_0\left(\cD{\ell}{1}\right) =
\lam^2-\left({d-1\ov d-1+2\ell}\right)^2\lam^2\;,\qquad \ell\in\N_0 \;.\]
The index $0$ at the energy $E$ indicates the $0$-particle sector.
Since the supercharges commute with the dynamical symmetry
group the multiplet $\cD\ell 1$ is paired with the 
same multiplet in $\cH_1$. The eigenfunctions in $\cH_1$ are obtained
by acting with $\Qd$ on those in $\cH_0$.
According to our previous results (see the figure above)
there exists the additional multiplet $\cD\ell 2$ in $\cH_1$.
This is obtained by acting with $Q$ on the same representation
in the two-particle sector. Hence $H=Q\Qd$ on this second
multiplet and we obtain
\begin{eqnarray*}
E_1\left(\cD{\ell}{1}\right) 
&=&\lam^2-\left({d-1\ov d-1+2\ell}\right)^2\lam^2\;, \qquad \ell\in\N \;, \\
E_1\left(\cD{\ell}{2}\right) &=&\lam^2-\left({d-3\ov
d-1+2\ell}\right)^2\lam^2\;, \qquad\ell\in \N\;.
\end{eqnarray*}
Note that $\ell=0$ does not occur in $\cH_1$. In $\cH_0$ the states
with $\ell=0$ have vanishing energy and hence are annihilated
by $\Qd$. 

Now one continues with $\cH_2$ and further on to
$\cH_3$ etc. One only needs the formulae
\begin{eqnarray*}
E_p(\cD\ell p)&=&
\Qd Q\Big|_{\cH_p}\left(\cD\ell p\right) 
=\lam^2-\left({d+1-2p \ov d-1+2\ell}\right)^2\lam^2\;,\qquad
\ell\in\N\;, \\
E_p(\cD\ell{p+1})&=&Q\Qd\Big|_{\cH_p}\left(\cD{\ell}{p+1}\right)
=\lam^2-\left({d-1-2p \ov d-1+2\ell}\right)^2\lam^2\;,
\qquad \ell\in\N\;.
\end{eqnarray*}
We shall determine the corresponding eigenfunctions
in the following section.

\section{Eigenstates of the supersymmetric hydrogen atom}
\label{sec6}
So far we have not considered which highest weight states of the
dynamical symmetry group are normalizable.
Now we explicitly construct these states in all subspaces
$\cH_p\subset \cH$. In the previous section we have seen
that for any $\ell\geq 1$ there are one or two irreducible
representations of $SO(d+1)$, namely
\eqnl{
\cD\ell p\subset \left(\cH_p\cap \Qd\cH\right)\mtxt{and}
\cD\ell{p+1}\subset \left(\cH_p\cap Q\cH\right)\;.}{andi1}
It suffices to construct the
highest weight states $\Psi_p(\cD\ell{p+1})$ 
of the latter multiplets. The highest weight states 
$\Psi_p(\cD\ell p)$ of the first set of multiplets in \refs{andi1}
are then just their superpartners,
\eqnn{
\Psi_p(\cD\ell p)=\Qd\Psi_{p-1}(\cD\ell p)\;.}
Actually we only need to calculate the
highest weight states $\Psi_p(\cD\ell{p+1})$ for 
$p<d/2$ because of the duality relation
\eqnn{
(p,\lam)\longleftrightarrow (d-p,-\lam)\;,}
which follows from \refs{sqm16} and \refs{lrl8}.

Observe that for any normalizable $H$-eigenstate 
$\Psi\in Q\cH$ the transformed state $\Qd \Psi$ 
is normalizable, as can be seen from 
\eqnn{
(\Qd\Psi,\Qd \Psi)=(\Psi, Q\Qd\Psi) =(\Psi,H\Psi)=E(\Psi,\Psi)\;.} 
Without calculating the highest weight
states we can argue in which sectors
bound states cannot exist. For that purpose
we consider the Hamiltonian \refs{lrl11}. 
It is easy to see that the hermitian
operator $S^\dagger S$, where $S$ has been defined in \refs{lrl5},
is an orthogonal projector, and hence has eigenvalues $0$ and $1$.
It follows at once that for $p>d/2$ the operator
$A$ in \refs{lrl10} is negative and hence $H>\lambda^2$.
We conclude that $H$ has no bound states 
in the sectors $\cH_{p>d/2}$.
On the particular sector $\cH_n$ 
the operator $A$ has both positive
and negative eigenvalues.
We expect that in this sector only one of the 
two representations (for each $\ell$) of the dynamical
symmetry group contains bound states. After
these general considerations we proceed with 
computing the highest weight states
$\Psi_p(\cD\ell{p+1})$ in the subspace $\cH_p\cap Q\cH$.
Again we proceed differently in even- and odd-dimensional
spaces.

\textbf{Even-dimensional spaces}: 
We use the complex coordinates $z_1,\ldots, z_n$ in
$\R^{d=2n}$ and the creation/annihilation operators
$\phid_1, \ldots, \phid_n, \phi_1, \ldots, \phi_n$
introduced in appendix \ref{append1a}, together
with the complex conjugated objects. 
Since the dynamical symmetry group is $SO(2n\ps 1)$,
we should take the Cartan operators $H_i$ and raising operators
$E_i$ from appendix \refs{append1b}. However, we must
remember that the last row and last column of $(J_{AB})$
contain the components of the generalized
Laplace-Runge-Lenz vector. As a consequence the
last step operator $E_n$ in \refs{apen24b} is 
to be replaced by
{\small
\eqngrl{
E_n&=&\has(K_{d-1}+\im K_d)\sim\has (C_{d-1} + \im C_d )}
{&=&-2z_n \Delta +(2r\pa_r + d - 1) \bar\pa_n
-2\phid_n\big(\phi_i\bar\pa_i +\bar\phi_i\pa_i\big)
+2\big(\phid_i\pa_i + \phib^\dagger_i \bar\pa_i\big)\bar{\phi}_n
-\frac{\lambda z_n}{r} A\;,}{expl1}}
Since the simple roots of the rotational subgroup $SO(d)$
are positive roots of the dynamical symmetry group, 
a highest weight state of $SO(d\ps 1)$
is automatically a highest weight state of $SO(d)$, similar as
in the purely bosonic case. Since the two groups
share the same Cartan generators the highest
weight state $\Psi_p(\cD\ell {p+1})$ of $SO(d\ps 1)$
must also be a highest weight state of the 
multiplet $\cD\ell{p+1}$ of $SO(d)$. From the branching rule
\refs{bra1} and the tensor products \refs{alg10} it 
follows, that this highest weight state must be the 
state $\cY_a(\ell,p+1)$ given in \refs{apen21}. 
Hence we are lead to the
ansatz
\eqn{
\Psi_p(\cD\ell{p+1}) = f(r) \cY_a(\ell,p+1) \;.}
It remains to determine the radial 
function $f(r)$ such that $\Psi_p$ is
annihilated by $E_n$. With $E_n$ from  \refs{expl1} one finds
the following equation for the radial function $f$:
\eqnn{
(d-1+2 \ell) f'+\lambda (d-1-2p)f= 0 }
such that the relevant highest weight states in the
$p$-particle take the form
\eqnl{
\Psi_p(\cD\ell{p+1})
= \exp\left(-\gamma_{\ell p} r\right)\, \cY_a(\ell,p+1)
\quad\mtxt{with}
\quad \gamma_{\ell p} ={d-1-2p\ov d-1+2\ell}\,\lam\;.}{andi2}
As $\lambda$ is assumed positive, these are bound states 
for $p< n$. 

\textbf{Odd-dimensional spaces:} 
For odd dimensions $d=2n\ps 1$ the rank of the dynamical
symmetry group $SO(2n\ps 2)$ exceeds the rank of the rotation group $SO(d)$
by one. As in the purely bosonic case we combine the 
first $2n$ coordinates, creation and annihilation
operators to complex ones (cf. appendix \ref{append1b}).
Since the rank of the dynamical symmetry group is
even, we should take the Cartan generators from
appendix \ref{append1a} with $n$ replaced by $n+1$.
Since $(J_{AB})$ contains the Laplace-Runge-Lenz
vector the explicit realization of these operators
differs from the one in this appendix. More precisely,
the first $n$ Cartan generators are those in \refs{apen7},
but the last one $H_{n+1}$ is $K_d\sim C_d$ (cf. appendix \ref{append2}), where
{\small
\eqnl{
C_d=-2 x_d \Delta +(2r\partial_r + d -1)\partial_d 
-2\psi^\dagger_d(\phi_i\bar\pa_i +\phib_i\pa_i) 
+2(\phid_i \pa_i + \phib^\dagger_i\bar\pa_i)
\psi_d - \lambda \hat{x}_d A \;.}{expl10}}
The raising operators are the $n - 1$ operators $E_i$ in 
\refs{apen9a} plus the two operators\footnote{which
are independent combinations of the last two
raising operators, see appendix \ref{append2}.}
\eqnl{
E_n=\frac{1}{\im}(z_n\pa_{x_d} - x_d \bar\pa_{z_n}
+\phid_n\psi_d-\psid_d\phib_n)\mtxt{and}
E_{n+1} = \has (C_{d-2} + \im C_{d-1})\;.}{expl11}
The last operator coincides with $E_n$ in \refs{expl1}.
By using similar arguments as in even dimensions we are
lead to the following ansatz
\eqnn{
\Psi_p(\cD\ell{p+1})= f(r)\, \cY_a(\ell,p+1) }
for the highest weight state of $\cD\ell{p+1}\subset\cH_p$.
This function is annihilated by all $E_{i\,\leq \,n}$. 
The condition $E_{n+1}\Psi_p=0$ yields the same differential 
equation for the radial function $f$ as before and we obtain
\eqnl{
\Psi_p(\cD\ell{p+1})
= \exp\left(-\gamma_{\ell p} r\right)\, \cY_a(\ell,p+1)
\quad\mtxt{with}\quad 
\gamma_{\ell p} ={d-1-2p\ov d-1+2\ell}\,\lam\;.}{expl13}
For positive $\lam$ these states are normalizable
for all $p<n$.
It is easy to see that the last Cartan generator
$\sim C_d$ annihilates this state and this
shows that it has the correct weight.

\textbf{The remaining highest weight states:}
We have argued that the highest weight
state $\Psi_{p+1}(\cD\ell{p+1})\subset \cH_{p+1}$
is the superpartner of
$\Psi_p(\cD\ell {p+1})$ in \refs{expl13}.
A simple calculation yields
\eqngrl{
\Psi_{p+1}(\cD\ell{p+1})&=&Q^\dagger\Psi_p(\cD\ell{p+1})}
{&=&\im\Big((\lam-\gam_{\ell p})S^\dagger\cY_a(\ell,p+1)
+(\ell+p)\cY_s(\ell,p+1)\Big)
\exp(-\gam_{\ell p}r)}{expl4}
for this state and shows that it is a 
linear combination of the two highest weight
states $\cY_s$ and $\cY_a$ of $SO(d)$
given in formulae (\ref{apen20},\ref{apen21}).
These states lead to additional 
bound-state multiplets in the sectors $\cH_p$
with $p=1,\dots,n$.

\section{The supersymmetric hydrogen atom in dimensions $\leq$ 4}
\label{sec7}
In this section we apply the general results
of the previous three sections to supersymmetric
systems in low dimensions. The two-dimensional
case mainly serves as illustration of the method.
It may be worth noting that such systems admit a supersymmetric Laplace-Runge-Lenz vector, contrary
to what has been claimed in the literature \cite{heumann}.
The three-dimensional quantum system is of course the 
most interesting extension of the ordinary hydrogen atom. 
We have included the four-dimensional
supersymmetric system since it already shows very
nicely which additional structures arise
when one goes beyond three dimensions.

\textbf{Two dimensions:}
The Hilbert space of the supersymmetric
hydrogen atom in two space dimensions contains
three sectors,
\[
\cH=\cH_0\oplus\cH_1\oplus \cH_2 \;,\]
and an arbitrary wave function has the expansion (in the basis \refs{states})
\eqnn{
\Psi=f_0\vac +(f_1\vert 1\ra+f_2\vert 2\ra)+f_{12}\vert 12\ra
\sim f=(f_0,f_1,f_2,f_{12})^T.}
The Hamiltonian \refs{lrl11} acts on 
the component functions in $f$ as follows
\[
f\longrightarrow Hf\mtxt{with}
H=-\triangle+\lam^2+{\lam \ov r}\pmatrix{-1&0&0\cr 0& \delta_{ab}-
2\hat x_a\hat x_b&0\cr 0&0&1}\]
Clearly, for $\lam>0$ there are no bound 
states in the two-particle subspace,
in accordance with our general result below
eq. \refs{andi1}. Only the multiplets
\eqnn{
\cD{\ell\,\geq\, 0} 1\subset \cH_0\mtxt{and}
\cD{\ell\,>\,0} 1\subset\cH_1}
of the dynamical symmetry group $SO(3)$ 
contain normalizable states. 

We introduce the complex
coordinate $z$ and the complex creation/annihilation
operator (see appendix \ref{append1a})
in terms of which the highest weight state
\refs{andi2} read
\[
\Psi_0(\cD\ell 1)=\exp(-\gam_{\ell 0}r)\cY_a(\ell,1)=
\exp(-\gam_{\ell 0}r)\,z^\ell\vac \;,\quad
\gam_{\ell 0}={\lam\ov 1+2\ell} \;,\]
and its superpartner \refs{expl4}
\eqnl{
\Psi_1(\cD\ell 1)=\Qd\Psi_0(\cD\ell 1)
=\im \Big((\lam\ms\gam_{\ell 0})S^\dagger\cY_a(\ell,1)+\ell \cY_s(\ell,1) 
\Big)\exp(-\gam_{\ell 0}r)\;.}{ex2d4}
The energy of the $\ell\ps 1$ states in each of the
two corresponding $SO(3)$-multiplets is
\[
E_\ell=\lam^2-\gam^2_{\ell 0} \;.\]
There is exactly one zero-energy ground state
in the zero-particle sector and this state
has $\ell=0$. It is annihilated by 
the adjoint supercharge $\Qd$, as can be seen
from \refs{ex2d4}. The spectrum of the supersymmetric
system is depicted in the following figure.

\setlength{\unitlength}{0,03mm}\label{figure1}
\begin{center}
\begin{picture}(2500,2150)(0,-200)
\multiput(200,0)(1000,0){3}{\line(0,1){300}}
\multiput(200,380)(1000,0){3}{\line(0,1){1250}}
\multiput(150,300)(1000,0){3}{\line(5,1){100}}
\multiput(150,350)(1000,0){3}{\line(5,1){100}}
\multiput(0,1600)(60,0){40}{\line(1,1){50}}
\put(150,1790){$\cH_0$}
\put(1150,1790){$\cH_1$}
\put(2150,1790){$\!\!\!\!\!\cH_2$}
\put(2200,1600){\vector(0,1){100}}
\put(2250,1700){\small $ E/\lam^2$}
\put(2450,150){\small $0$}
\put(2450,450){${8\ov 9}$}
\put(2450,1550){\small $1$}
\dottedline{40}(0,200)(2400,200)
\dottedline{40}(0,1600)(2400,1600)
\dottedline{40}(0,500)(2400,500)
\put(0,200){\line(1,0){400}}
\multiput(0,500)(1000,0){2}{\line(1,0){400}}
\multiput(0,1204)(1000,0){2}{\line(1,0){400}}
\multiput(0,1398)(1000,0){2}{\line(1,0){400}}
\multiput(0,1478)(1000,0){2}{\line(1,0){400}}
\multiput(0,1518)(1000,0){2}{\line(1,0){400}}
\multiput(0,1541)(1000,0){2}{\line(1,0){400}}
\multiput(0,1556)(1000,0){2}{\line(1,0){400}}

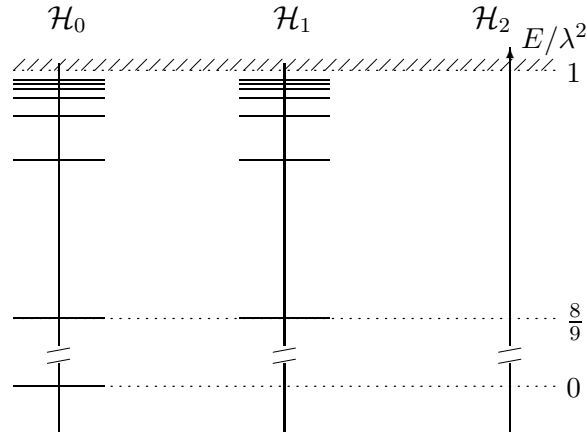
\put(100,-200){Eigenvalues of $H$ in $d=2$ dimensions.}
\end{picture}
\end{center}

\textbf{Three dimensions:}
For this most relevant system, the
Hilbert space contains four sectors,
\[
\cH=\cH_0\oplus \cH_1\oplus\cH_2\oplus\cH_3 \;,\]
and a wave function has the expansion
\[
\Psi=f_0\vac+\big(f_1\vert 1\ra+f_2\vert 2\ra
+f_3\vert 3\ra\big)+\dots \;.\]
Since here we are only interested in bound states
it suffices to consider $H$
on the subsectors $\cH_0$ and $\cH_1$. The
Hamiltonian in $\cH_0$ belongs to the
ordinary hydrogen atom,
\[
H^{(0)}=-\triangle +\lam^2-{2\lam\ov r} \;,\]
and it acts on a state $\Psi\in\cH_1$ with
component functions $\la a\vert H\ra=f_a$ as follows,
\[
\la a\vert H\Psi\ra=\big(-\triangle+\lam^2)f_a
-{2\lam\ov r}\hat x_a \hat x_b f_b \;.\]
We take the coordinates $(z,x_3)$ and the creation
operators $(\phid,\phib^\dagger,\psid_3)$ (see 
appendix \ref{append1b}).
The highest weight state in the
multiplet $\cD\ell 1\subset \cH_0$ 
of the dynamical symmetry group $SO(4)$ has the form
\[
\Psi_0(\cD\ell 1)=\exp(-\gam_{\ell 0}r)\cY_a(\ell,1)\in \cH_0 \;,\quad
\gam_{\ell 0}={\lam\ov 1+\ell}\;,\]
and this state is mapped into the partner state
\[
\Psi_1(\cD\ell 1)=\Qd\Psi_0(\cD\ell 1)
=\im\Big( (\lam-\gam_{\ell 0})S^\dagger\cY_a(\ell,1)
+\ell \cY_s(\ell,1) 
\Big)\exp(-\gam_{\ell 0}r) \;.\]
All states in the two corresponding $SO(4)$-multiplets 
share the same energy
\[ E_\ell=\lam^2-\gam_{\ell 0}^2 \]
and both multiplets contain $(\ell+1)^2$ states.
This is the well-known spectrum of the hydrogen
atom. The normalizable zero-mode has $\ell=0$ and resides
in the zero-particle sector. It is just the ground
state of the hydrogen atom.

There are no further bound states, since
the other states are paired
with wave functions in $\cH_2$ and thus cannot be normalizable. 
The spectrum of the supersymmetric hydrogen atom
in three dimensions is shown below.

\setlength{\unitlength}{0,026mm}\label{figure2}
\begin{center}
\begin{picture}(3500,2150)(0,-200)

\multiput(200,0)(1000,0){4}{\line(0,1){300}}
\multiput(200,380)(1000,0){4}{\line(0,1){1250}}
\multiput(150,300)(1000,0){4}{\line(5,1){100}}
\multiput(150,350)(1000,0){4}{\line(5,1){100}}
\multiput(0,1600)(60,0){56}{\line(1,1){50}}
\put(150,1790){$\cH_0$}
\put(1150,1790){$\cH_1$}
\put(2150,1790){$\cH_2$}

\put(3150,1790){$\!\!\!\!\!\cH_3$}
\put(3200,1600){\vector(0,1){100}}
\put(3250,1700){\small $ E/\lam^2$}
\put(3450,450){${3\ov 4}$}
\put(3450,150){\small $0$}
\put(3450,1550){\small $1$}
\dottedline{40}(0,200)(3400,200)
\dottedline{40}(0,1600)(3400,1600)
\dottedline{40}(0,500)(3400,500)
\put(0,200){\line(1,0){400}}
\multiput(0,500)(1000,0){2}{\line(1,0){400}}
\multiput(0,1111)(1000,0){2}{\line(1,0){400}}
\multiput(0,1325)(1000,0){2}{\line(1,0){400}}
\multiput(0,1424)(1000,0){2}{\line(1,0){400}}
\multiput(0,1478)(1000,0){2}{\line(1,0){400}}
\multiput(0,1510)(1000,0){2}{\line(1,0){400}}
\multiput(0,1531)(1000,0){2}{\line(1,0){400}}
\multiput(0,1546)(1000,0){2}{\line(1,0){400}}
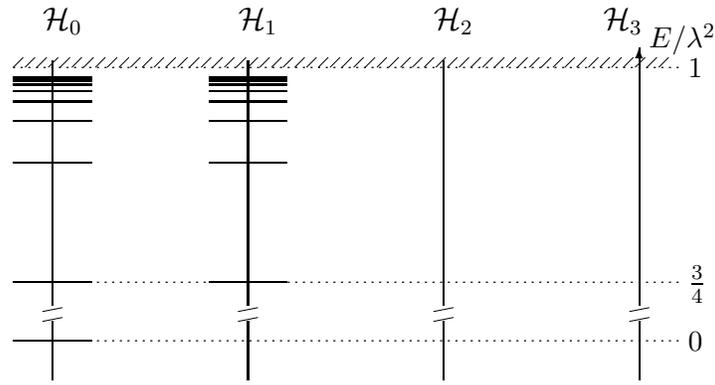
\put(400,-200){Eigenvalues of $H$ in $d=3$ dimensions.}
\end{picture}
\end{center}

\textbf{Four dimensions:}
The Hilbert space splits into five subsectors,
\[
\cH=\cH_0\oplus\cH_1\oplus \cH_2\oplus\cH_3\oplus\cH_4 \;.\]
In the zero-particle sector the Hamiltonian
takes the form,
\[
H^{(0)}=-\triangle +\lam^2-{3\lam\ov r} \;,\]
and in the one-particle sector it acts on 
a state $\Psi$ with $\la a\vert\Psi\ra=f_a$ as
follows:
\[
\la a\vert H\Psi\ra=
\left(-\triangle+\lam^2-{\lam\ov r}\right)f_a
-{2\lam \ov r}\hat x_a\hat x_b f_b\;.\]
In $\cH_2$ we are left with a $6\times 6$-matrix 
Schr\"odinger operator: For a two-particle state
$\Psi$ with component function 
$\la ab\vert \Psi\ra=f_{ab}$ we obtain
\[
\langle ab\vert H\Psi\ra=\left(-\triangle+\lam^2
+{\lam\ov r}\right)f_{ab}-{2\lam\ov r}\big(
\hat x_b f_{ac}-\hat x_a f_{bc}\big)\hat x_c\;.\]
We introduce complex coordinates $z_1,z_2$ and 
creation operators $\phid_1,\phid_2$ as in appendix \ref{append1a}.
For all $\ell\in \N$ we find the highest weight states
\[
\Psi_0(\cD\ell 1)= \exp(-\gam_{\ell 0}r)\,z_1^\ell\, \vac
\mtxt{with}
\gamma_{\ell 0}={3\lam\ov 2\ell+3}\]
in the zero-particle sector, together with their 
superpartners in $\cH_1$,
\[
\Psi_1(\cD\ell 1)=
\im\Big((\lam-\gam_{\ell 0})S^\dagger\cY_a(\ell,1)
+\ell \cY_s(\ell,1) 
\Big)\exp(-\gam_{\ell 0}r)\;,\]
which exists for $\ell>0$. All states in these two
multiplets have the same energy
\[
E_\ell=\lam^2-\gam_{\ell 0}^2 \;,\]
and the number of states in each multiplet is
$(\ell+1)(\ell+2)(2\ell+3)/6$.
Now there is an additional representation
$\cD\ell 2\subset\cH_1$ with highest weight state
\eqngr{
\Psi_1(\cD\ell 2)&=&\exp(-\gam_{\ell 1}r)\cY_a(\ell,2)}
{&=&\exp(-\gam_{\ell 1}r)\,
\big(\phid_2z_1-\phid_1 z_2)\,z_1^{\ell-1}\vac\;,\qquad
\gam_{\ell 1}={\lam\ov 3+2\ell}\;.}
The energy of this state is
\[
E_\ell=\lam^2 -\gam_{\ell 1}^2\;,\]
and the multiplet contains
$\ell (\ell+3)(2\ell+3)/4$ members.
Again there is a supersymmetric
partner multiplet
with the same energy and degeneracy. The remaining highest 
weight state of $\cD\ell 3 \simeq \cD\ell 2\subset\cH_2$
is paired to a state in $\cH_3$ and therefore not normalizable. 
Thus we find the spectrum as depicted in the following
figure.

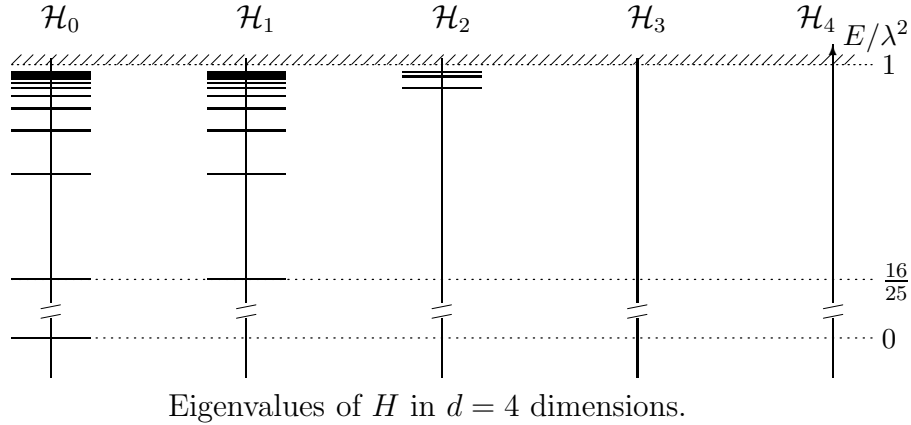
\begin{center}
\begin{picture}(4500,2150)(0,-200)

\multiput(200,0)(1000,0){5}{\line(0,1){300}}
\multiput(200,380)(1000,0){5}{\line(0,1){1250}}
\multiput(150,300)(1000,0){5}{\line(5,1){100}}
\multiput(150,350)(1000,0){5}{\line(5,1){100}}
\multiput(0,1600)(60,0){72}{\line(1,1){50}}
\put(150,1790){$\cH_0$}
\put(1150,1790){$\cH_1$}
\put(2150,1790){$\cH_2$}
\put(3150,1790){$\cH_3$}
\put(4150,1790){$\!\!\!\!\!\cH_4$}
\put(4200,1600){\vector(0,1){100}}
\put(4250,1700){\small $ E/\lam^2$}
\put(4450,450){${16\ov 25}$}
\put(4450,150){\small $0$}
\put(4450,1550){\small $1$}
\dottedline{40}(0,200)(4400,200)
\dottedline{30}(0,1600)(4400,1600)
\dottedline{40}(0,500)(4400,500)
\put(0,200){\line(1,0){400}}
\multiput(0,500)(1000,0){2}{\line(1,0){400}}
\multiput(0,1039)(1000,0){2}{\line(1,0){400}}
\multiput(0,1260)(1000,0){2}{\line(1,0){400}}
\multiput(0,1373)(1000,0){2}{\line(1,0){400}}
\multiput(0,1437)(1000,0){2}{\line(1,0){400}}
\multiput(0,1478)(1000,0){2}{\line(1,0){400}}
\multiput(0,1504)(1000,0){2}{\line(1,0){400}}
\multiput(0,1524)(1000,0){2}{\line(1,0){400}}
\multiput(0,1538)(1000,0){2}{\line(1,0){400}}
\multiput(0,1548)(1000,0){2}{\line(1,0){400}}
\multiput(0,1556)(1000,0){2}{\line(1,0){400}}
\multiput(0,1562)(1000,0){2}{\line(1,0){400}}
\multiput(1000,1477)(1000,0){2}{\line(1,0){400}}
\multiput(1000,1538)(1000,0){2}{\line(1,0){400}}
\multiput(1000,1562)(1000,0){2}{\line(1,0){400}}
\put(800,-200){Eigenvalues of $H$ in $d=4$ dimensions.}
\end{picture}
\end{center}

\section{Summary and Discussion}
\label{sec8}
We have succeeded in extending the celebrated
results of Pauli, Fock, Bargmann and others in two
directions: to higher dimensions and to
the $\cN=2$ supersymmetric extension of the hydrogen
atom. First we constructed a generalized 
angular momentum and an extended
Laplace-Runge-Lenz vector which 
could be combined to generators
of the dynamical symmetry group $SO(d\ps 1)$
in $d$ dimensions. Then we related the quadratic Casimir 
operator of this group to the particle number $N$, $Q\Qd$ and
$Q\Qd$. This way we calculated the
bound state spectrum of the supersymmetric hydrogen atom
in arbitrary dimensions by algebraic means.
We have determined all bound-state multiplets
of the dynamical symmetry group
and calculated the highest weight state in each
of them.

Itzykson and Bander \cite{bander} distinguished
between the infinitesimal and the global
method to solve the Coulomb problem.
The former is based on the Laplace-Runge-Lenz
vector and is the method used in this paper.
In the second method one performs a stereographic
projection of the $d$-dimensional momentum
space to the unit sphere in $d\ps 1$ dimensions
which in turn implies a $SO(d\ps 1)$
symmetry group. It would be interesting to
perform a similar global construction for
the supersymmetric systems introduced in
this paper.

We have not explained why every multiplet
of the dynamical symmetry group appears four
times. Furthermore there is a new 'accidental' degeneracy: in higher
dimensions some eigenvalues of
the Hamiltonian appear in many different particle-number sectors. It may very
well be, that the algebraic
structures discussed in the present 
work have a more natural setting in the language 
of superalgebras. We have not
investigated this question.
Anonther interesting question is whether the dynamical symmetries considered in
this paper are related to fermionic Killing-Yano supercharges \cite{holten}.
This problem needs further investigations.
Finally, we see no obstacle against extending our
method to the scattering
problem of the supersymmetric hydrogen atom,
for which the dynamical symmetry group 
$SO(1,d)$ is non-compact.

\begin{acknowledgments} We thank Falk Bruckmann 
for his collaboration at an early stage of
this project and Thomas Strobl and Jan-Willem van Holten for useful
discussions. P.A.G. Pisani thanks people at the TPI for their kind hospitality.
A. Kirchberg and P.A.G. Pisani are supported by the Studienstiftung des
Deutschen Volkes and CONICET, respectively. This project has in
part been supported
by Foundaci\'on Antorchas and DAAD (grant 13887/1-87). Most group-theoretical
calculations have been performed with the powerful
package LiE \cite{lie}.

\end{acknowledgments}

\begin{appendix}
\section{Representations of rotation groups}
\label{append1}

In this appendix we collect the group theoretical
facts needed in the main body of the paper (cf. \cite{hamermesh, humphreys, mckay, varadarajan} for a more detailed discussion of these issues). 
We shall construct the relevant irreducible representations of the
total angular momentum operators
\[
J_{ab}={1\ov \im}\left(x_a{\pa\ov \pa x_b}
-x_b{\pa\ov\pa x_a}\right)
+{1\ov \im}\left(\psid_a\psi_b-\psid_b\psi_a\right),\qquad
a,b=1,\dots,d \]
satisfying the $so(d)$ commutation relations
\[
[ J_{ab}, J_{cd} ] = \im (\delta_{ac} J_{bd} 
+ \delta_{bd} J_{ac} - \delta_{ad} J_{bc} - \delta_{bc}
J_{ad} ) \]
on wave functions in 
\eqnn{
\cH_p=
L_2(\R^d)\times \C^{d\choose p}\mtxt{with}p=0,\dots,d\;.}
The fermionic operators $\psi_a$ have been introduced
earlier in section \ref{sec3}.
It is convenient to use the 
Cartan-Weyl basis consisting of generators $H_i$ in 
the Cartan subalgebra and one raising 
and one lowering operator $E_\al$ and $E_{-\al}$
for every positive root $\al$,
\eqnl{
[H_i,E_\al]=\al_i E_\al\mtxt{and}
[E_\al,E_{-\al}]=\al\cdot H\mtxt{with} 
E_{-\al}=E^\dagger_\al\;.}{apen3}
Because of their different properties we do this
separately for the groups $D_n\sim SO(2n)$ 
and $B_n\sim SO(2n\ps 1)$.

\subsection{Total angular momentum for the $SO(d=2n)$ groups}
\label{append1a}
To proceed it is very convenient to introduce
complex coordinates in $\R^{2n}$,
\eqngrl{
z_i = \has (x_{2i-1}+\im x_{2i})\;, && \bar z_i=\has (x_{2i-1}-\im
x_{2i})\;,}
{\pa_i  = \has
(\pa_{x_{2i-1}}-\im\pa_{x_{2i}})\;, && \bar\pa_i=\has(\pa_{x_{2i-1}} +
\im\pa_{x_{2i}})\;, \quad i = 1, \ldots n} {apen4}
and similarly two sets of complex creation- and annihilation operators
\eqngrl{
\phid_i = \has (\psid_{2i-1}+\im\psid_{2i})\;,&&
\phib^\dagger_i=\has (\psid_{2i-1} - \im\psid_{2i})\;,}
{\phi_i =\has
(\psi_{2i-1}-\im\psi_{2i})\;, &&
\phib_i=\has(\psi_{2i-1}+\im\psi_{2i})\;, \quad i = 1, \ldots n \;.}
{apen5}
The only non-vanishing anticommutators are
\[
\{\phi_i,\phid_j\}=
\{\phib_i,\phib_j^\dagger\}=\delta_{ij}\;.\]
The generators in the Cartan subalgebra take the simple form
\eqnl{
H_i=J_{2i-1,2i}=z_i\pa_i-\bar z_i\bar\pa_i+\phid_i\phi_i
-\phib^\dagger_i\phib_i\;,
\qquad i=1,\dots,n}{apen7}
and there are two types of raising operators:
\eqngr{
 E_\al&=&\ha\big(J_{2i-1,2j-1}+J_{2i,2j}-\im J_{2i-1,2j}+ \im J_{2i,2j-1})
\mtxt{with root}\al=e_i-e_j\;,}
{E_\al&=&\ha\big(J_{2i-1,2j-1}-J_{2i,2j}+ \im J_{2i-1,2j}+\im J_{2i,2j-1})
\mtxt{with root}\al=e_i+e_j\;,}
where $i<j$ is assumed. In terms of the complex
coordinates/operators
they read
\eqngrl{
E_\al
&=&{1\ov \im}\left(z_i\pa_j-\bar z_j\bar\pa_i
+\phid_i\phi_j-\phib^\dagger_j\phib_i\right)\mtxt{with root}\al=e_i-e_j\;,}
{E_\al
&=&{1\ov \im}\left(z_i\bar\pa_j-z_j\bar\pa_i+
\phid_i\phib_j-\phid_j\phib_i\right)\mtxt{with root}\al=e_i+e_j\;.}{apen8}
The corresponding lowering operators
are just the adjoint of the raising operators.
The operators ($H_i,E_\al,E_{-\al}$) satisfy the
commutation relations \refs{apen3} with corresponding
positive roots in \refs{apen8}.
The $n$ simple roots are
\eqnn{
\al_i=e_i-e_{i+1}\;,\quad 1\leq i< n\quad\mtxt{and}\quad 
\al_n=e_{n-1}+e_n}
and the corresponding raising operators have the form
\begin{eqnarray}
E_i&=&{1\ov \im}\left(z_i\pa_{i+1}-\bar z_{i+1}\bar\pa_i
+\phid_i\phi_{i+1}-\phib^\dagger_{i+1}\phib_i\right),\quad
\al=e_i-e_{i+1},\;1\leq i<n\label{apen9a}\\
E_n&=&{1\ov \im}\left(z_{n-1}\bar\pa_n-z_n\bar\pa_{n-1}
+\phid_{n-1}\phib_n-\phid_n\phib_{n-1}\right),\quad
\al=e_{n-1}+e_n\;.\label{apen9b}
\end{eqnarray}
With the help of the Weyl vector
\[
\delta=\ha \sum_{\al >0}\al=(n-1)e_1+(n-2)e_2+\dots+ e_{n-1}\;,\]
where the sum extends over all positive roots in \refs{apen8},
we may calculate the dimension of an arbitrary
faithful representation of $SO(2n)$. Such a representation
is determined by its Young tableau containing at most
$n$ rows. The length $\ell_i$ of row $i$
is bigger or equal to that of row $i+1$. Hence, a
Young tableau is given by $n$ non-negative ordered integers
\def\ello{\ell_1}
\def\ellt{\ell_2}
\def\elln{\ell_n}
\def\ellp{\ell_p}
\eqnn{
\ello\geq \ellt\geq\dots\geq\ell_{n-1}\geq \ell_n} 
and has the form
\eqnn{
p\left\{\young(12\cdot\cdot\cdot
\ello,12\cdot\cdot\ellt,::::,1\cdot\ellp)\right.\;, \quad p \leq n\;.}
Rows with length $0$ are not shown when one
draws a Young-tableau.
The corresponding representation ${\cal D}^{\,\ell_1,\dots,\ell_n}$
has the dimension
\eqnl{
\hbox{dim}\left({\cal D}^{\,\ell_1\dots\ell_n}\right)=
\prod_{1\leq r<s\leq n}{\ell_r+ \ell_s+ 2n- r- s\ov 2n- r-s}\;
{\ell_r-\ell_s+s-r\ov s-r} \;.}{apen11}
For the second-order Casimir invariant of these
representations one obtains the formula
\[
\mathcal{C}_{(2)}({\cal D}^{\,\ell_1\dots\ell_n})=
\sum_r \ell_r (\ell_r + 2n - 2r) \;.\]
In particular, for the completely symmetric representations
\[
{\cal D}^{\,\ell 0\dots 0}\equiv \cD\ell 1\sim
\young(12\cdot\cdot\ell)\]
these formulae simplify to
\eqnl{
\cC_{(2)}(\cD\ell 1)=\ell(\ell+d-2)\mtxt{and}
\hbox{dim}(\cD\ell 1)={\ell\ps
d\ms 1\choose\ell}-{\ell\ps d\ms 3\choose \ell\ms 2}\,.}{apen14}
For the completely antisymmetric representations
\[
{\cal D}^{1,1,\dots 1}\equiv \cD 1p\sim
\young(1,:,p)\]
they simplify to
\[
\cC_{(2)}(\cD1p)=p(d-p)\mtxt{and}
\hbox{dim}(\cD1p)={d\choose p}\;.\]
Simultaneous eigenstates of all $n$ generators $H_i$
in the Cartan subalgebra have the form
\eqnn{
\prod_{i=1}^n z_i^{m_i}\bar z_i^{\bar m_i}\ket{\vec{p}\,
\vec{p}\,'}\;,\qquad
\ket{\vec{p}\,\vec{p}\,'}
=\phi_1^{\dagger\,p_1}\ldots\phi_n^{\dagger\,p_n}
\phib_1^{\dagger\,p_1'}\ldots \phib_n^{\dagger\,p_n'}
\vac\;,}
where $m_i,\bar m_i\in \N_0$ and $p_i,p_i'\in \{0,1\}$.
The vacuum $\vac$ is
annihilated by all particle lowering
operators $\psia$ or equivalently by all
$\phi_i$ and $\phib_i$. The $H_i$-eigenvalues
of these states are $m_i\ms\bar m_i\ps p_i\ms p_i'$.

Next we must construct the highest weight states 
which are annihilated by all raising operators. Every such
state determines an irreducible representation.
The eigenvalues of $H_i$  on a highest weight state is
equal to the length $\ell_i$ of the
Young tableau corresponding to the
irreducible representation determined by 
this weight. The $d+2$ space-independent highest weight states are
\begin{eqnarray*}
\vert p\ra= \vert \vec{p}\,\vec{p}\,'\ra &&\mtxt{with}
p_1\geq \dots \geq p_n\geq\, p_n'\geq\dots\geq p_1' \\
&&\mtxt{and} \sum (p_i+p_i')=p\;,
\end{eqnarray*}
There is an additional highest weight state in the $p=n$ particle sector, that arises since in this
sector we have selfdual and anti-selfdual configurations. It is given by
\[p_1 = \ldots = p_{n-1} = p_n' =1 \;, \qquad p_n = p_1' = \ldots = p_{n-1}'=0\;.\]
Clearly, the particle number $p$ uniquely
specifies these state since the $p_i, p_i'$'s are ordered.
These states define the completely antisymmetric
representations
\eqnn{
\cD1{p}\mtxt{for} p\leq n\quad\mtxt{and}
\cD1p\sim\cD1{2n-p}\mtxt{for}p\geq n\;.}
We used that a Young tableau, the first 
column of which has length $n\leq p\leq 2n$,
gives rise to the same multiplet as the 
tableau with first column of length $2n\ms p\leq n$.
It the following one should replace
$\cD\ell p$ by $\cD\ell{2n-p}$ if $p$ exceeds $n$.
Also note that $\cD10\sim\cD1d$ is the one-dimensional
trivial representation. 

The highest weight states in the $0$-particle sector
are
\eqnl{
z_1^{\ell}\vac}{apen18}
and they give rise to the completely symmetric
representations $\cD\ell 1$ spanned by the
harmonic polynomials of order $\ell$.
The relevant irreducible representation of $SO(2n)$
in the $p$-particle sector is gotten by
tensoring the antisymmetric representation
$\cD1p$ with a symmetric representation $\cD\ell 1$.
We use
\eqnl{
\cD1p\otimes \cD\ell 1=\cD\ell{p-1}\oplus \cD{\ell-1}p
\oplus\cD{\ell+1}p\oplus\cD\ell{p+1}}{apen19}
or in the language of Young tableaux,
\[
\young(1,:,:,p)\otimes
\young(12\cdot \ell)=
p\ms 1\left\{\young(12\cdot \ell,:,:)\right.\oplus
\overbrace{\young(1\cdot\cdot,:,:,p)}^{\ell-1}\oplus\;
\overbrace{\young(12\cdot\cdot\cdot,:,:,p)}^{\ell+1}\oplus\;
\left.\young(12\cdot \ell,:,:,:,:)\right\}p\ps 1\; .\]
Note that for $p=1$ and/or $\ell=1$ there appear 
only three representations in this decomposition. 
For $p=1$ the first representation and for $\ell=1$ the
second representation on the right hand side in
\refs{apen19} are absent. Also note that for $p=n$ the first
and last representations are equivalent.
The second to last representation $\cD{\ell+1}p$ 
on the right hand side has highest weight state
\eqnl{
\cY_s(\ell+1,p)=z_1^\ell\,\vert p\ra \;,}{apen20}
as it is the product of the highest weight states of $\cD{1}{p}$ and $\cD{\ell}{1}$.
To find the highest weight state of the other
representations we observe that the operators
\eqnn{
rS=x_a\psi_a= z_i\phi_i+\bar z_i\phib_i\mtxt{and}
rS^\dagger=x_a\psid_a=
\bar z_i\phid_i+z_i\phib^\dagger_i \;,}
which have been introduced in \refs{lrl5},
commute with the total angular momentum and hence
map highest weight states into highest weight states.
Since $S$ decreases and $S^\dagger$ increases the
particle number by one, we find the state
\eqnl{
\cY_a(\ell,p+1)&=&
rS\cY_s(\ell,p\ps 1)=\sum_{i=1}^{p+1} (-)^{i+1} z_i\phid_1\ldots\check\phi_i^\dagger
\ldots \phid_{p+1}z_1^{\ell-1}\vac}{apen21}
which is highest weight state of the last representation $\cD\ell{p+1}$ in the decomposition
\refs{apen19}. The missing two highest weight
states correspond to those representations in the
tensor product of a symmetric and an antisymmetric
representation which one obtains by taking the trace over
two suitable indices. This operation is equivalent to
acting with $S^\dagger$. Thus
\[\cT_s(\ell,p\ms 1)=S^\dagger\cY_s(\ell,p\ms 1)\]
is the highest weight state of $\cD\ell{p-1}$
in the decomposition \refs{apen19}. For the remaining highest 
weight state we make the ansatz
\eqnn{
\cT_a(\ell\ms 1,p)= \big(S S^\dagger
+ \alpha S^\dagger S\big) \cY_s(\ell\ms 1,p)\;.}
As $\{S,S^\dagger\}=1$ this state may have a 
component in the direction of $\cY_s(\ell\ms1,p)$.
However, for the choice $\alpha=-1$ the highest weight state
\[
\cT_a(\ell\ms 1,p)= [S,S^\dagger]\, \cY_s(\ell\ms 1,p)\;,\]
is orthogonal to $\cY_s(\ell\ms 1,p)$.

\subsection{Total angular momentum for the $SO(d=2n+1)$ groups}
\label{append1b}
The rotation group $SO(2n\ps 1)$ has the same
rank as its subgroup $SO(2n)$ and hence we may
still use the Cartan generators \refs{apen7},
that is
\eqnl{
H_i=J_{2i-1,2i}=z_i\pa_i-\bar z_i\bar\pa_i+\phid_i\phi_i
-\phib^\dagger_i\phib_i \;,
\qquad i=1,\dots,n \;.}{apen23a}
We use the complex coordinates \refs{apen4}
and the complex creation- and annihilation
operators \refs{apen5}, supplemented by
the last coordinate $x_d$ and the last creation
and annihilation operator $\psid_d$ and $\psi_d$.
Clearly, the raising operators \refs{apen8} are
still raising operators of $so(2n+1)$ with the
same positive roots. But since 
\[
\hbox{dim}\left(SO(2n\ps 1)\right)=
\hbox{dim}\left(SO(2n)\right)+2n\mtxt{and}
\hbox{rank}\left(SO(2n+1)\right)=
\hbox{rank}\left(SO(2n)\right)\]
there are $n$ positive roots missing. These are
\[
E_\al=\has \left(J_{2i-1,d}+\im J_{2i,d}\right)
={1\ov \im}\left(z_i\pa_{x_d}-x_d\bar\pa_i
+\phid_i\psi_d-\psid_d\phib_i
\right)\;,\quad
\al=e_i\;,\]
where $1\leq i\leq n$. The first $n\ms 1$ simple
roots are the same as in \refs{apen9a}, but the
the last one is replaced by $e_n$. Hence the
raising operators corresponding to the simple roots
read
\begin{eqnarray}
E_i&=&{1\ov \im}\left(z_i\pa_{i+1}-\bar z_{i+1}\bar\pa_i
+\phid_i\phi_{i+1}-\phib^\dagger_{i+1}\phib_i\right)\;,\quad\,
\al=e_i-e_{i+1}\;,\;\;1\leq i<n \;,\label{apen24a}\\
E_n&=&{1\ov \im}\left(z_n\pa_{x_d}-x_d\bar\pa_n
+\phid_n\psi_d-\psid_d\phib_n\right)\;,\quad\qquad \al=e_n\;.\label{apen24b}
\end{eqnarray}
The Young tableaux are
identical to those of $SO(2n)$ and hence
are characterized by $n$ ordered non-negative
integers $\ell_1,\dots,\ell_n$. The dimensions of
the corresponding representations read
\eqnl{
\hbox{dim}\left({\cal D}^{\,\ell_1\dots\ell_n}\right)=
\prod_{t=1}^n {2\ell_t+d-2t\ov d-2t}
\prod_{1\leq r<s\leq n}{\ell_r+\ell_s+d-r-s\ov d-r-s}\;
{\ell_r-\ell_s+s-r\ov s-r} }{apen25}
and the formula for the second-order Casimir
is the same as for the $so(2n)$ algebra,
\[
C_{(2)}({\cal D}^{\,\ell_1\dots\ell_n})=
\sum_r \ell_r (\ell_r + d - 2r) \;.\]
Also the rules for tensor products are identical to
those of $SO(2n)$.

Since the simple roots are different, the highest
weight states have a slightly different form.
The simultaneous eigenstates of the $n$ generators
in the Cartan subalgebra read
\eqnn{
f(x_d)\prod_i z_i^{m_i}\bar z_i^{\bar m_i}\,\ket
{\vec{p}\, q\,\vec{p}\,'}\; ,\qquad
\vert\vec{p}\, q\,\vec{p}\,'\ra
=\phi_1^{\dagger\,p_1}\ldots\phi_n^{\dagger\,p_n}\psi_d^{\dagger\,q}
\phib_1^{\dagger\,p_1'}\ldots \phib_n^{\dagger\,p_n'}
\vac \;,}
where $m_i,\bar m_i\in \Z$ and $p_i,q,\bar p_i\in \{0,1\}$.
The $d\ps 1$ constant highest weight states are
\[
\vert p\ra=\vert \vec{p}\, q\,\vec{p}\,'\ra\;\mtxt{with}
p_1\geq \dots\geq p_n\geq q\geq p_n'\geq\dots\geq p_1'\;,\]
where $p=\sum(p_i\ps p_i')+q$ denotes the particle number.
The highest weight of $\cD{\ell+1}p$ in the
decomposition
\eqnl{
\cD1p\otimes \cD\ell 1=\cD\ell{p-1}\oplus \cD{\ell-1}p
\oplus\cD{\ell+1}p\oplus\cD\ell{p+1}}{apen28}
is again determined by the highest weight state
\eqnn{
\cY_s(\ell+1,p)=z_1^l\,\vert p\ra \;.}
As in even dimensions one may use the scalar operators
\eqnn{
rS=x_a\psia=z_i\phi_i+\bar z_i\phib_i+x_d\psi_d\mtxt{and}
rS^\dagger=x_a\psid_a=\bar z_i\phid_i+z_i\phib^\dagger_i+x_d\psid_d}
to obtain the highest weight states 
\begin{eqnarray*}
\cY_a(\ell,p+1)&=& rS\cY_s(\ell,p\ps 1)\quad\longrightarrow\quad\cD\ell{p+1}\;, \\
\cT_s(\ell,p\ms 1)&=&S^\dagger\cY_s(\ell,p\ms 1)\quad\longrightarrow\quad\cD\ell {p-1}\;, \\
\cT_a(\ell\ms 1,p)&=&[S,S^\dagger]\,\cY_s(\ell\ms1,p)\quad\longrightarrow\quad\cD{\ell-1}{p}\;,
\end{eqnarray*}
of the remaining irreducible representations in \refs{apen28}.

\section{Rotation groups vs. dynamical symmetry groups}
\label{append2}
In the main body of the paper we have seen, that
the total angular momentum $J_{ab}$ in
\refs{lrl3} together with $K_a$
in \refs{alg2}
combine to generators of the dynamical symmetry
group $SO(d\ps 1)$
\[
J_{AB}=
\left(\begin{array}{c|c} J_{ab} & K_a \\ \hline 
-K_b & 0 \end{array}\right)\;.\]
The rotational group with generators $J_{ab}$
discussed in the previous part of the appendix,
must be embedded into the dynamical group,
\begin{eqnarray*}
d&=&2n:\qquad\;\, SO(2n)\subset SO(2n+1) \\
d&=&2n\ps 1:\quad SO(2n+1)\subset SO(2n+2)\;.
\end{eqnarray*}
\textbf{Even dimensions:}
The dynamical symmetry group has the same rank
as the rotation group $SO(2n)$ and we can
repeat our construction in appendix \ref{append1b},
where we extended $SO(2n)$ to $SO(2n\ps 1)$.
Of course we should take into account that the
components in the last column and last row
of $(J_{AB})$ are the components of $K_a$.
The Cartan generators are those in \refs{apen23a}
and the first $n\ms 1$ raising operators 
are given in \refs{apen24a}. But the last raising
operator \refs{apen24b} is of course replaced by
\[
E_n=\has (K_{d-1} + \im K_d )\;,\]
which is proportional to $\has (C_{d-1}+\im C_d)$. The
latter has been given in \refs{expl1}.

\textbf{Odd dimensions:}
The rank of the dynamical symmetry group $SO(2n\ps 2)$
exceeds the rank of the rotation group $SO(2n\ps 1)$ by
one. The Cartan generators are given by the $n$ operators
$H_i$ in \refs{apen7}, supplemented by $H_{n+1}=K_d\sim C_d$, 
where the explicit realization of $C_d$ is given in
\refs{expl10}.
The raising operators are the $n-1$ operators $E_i$ in 
\refs{apen24a} plus the two operators
\begin{eqnarray*}
E_\al&=&\ha\big(J_{d-2,d}+K_{d-1}-\im K_{d-2}+\im J_{d-1,d}\big)\;,
\quad \al=e_n-e_{n+1}\;, \\
E'_\al&=&\ha\big(J_{d-2,d}-K_{d-1}+\im K_{d-2}+\im J_{d-1,d}\big)\;,
\quad \al=e_n+e_{n+1}\;.
\end{eqnarray*}
Highest weight states are annihilated by these
two raising operators and it is convenient to
use two (independent) combinations of these 
operators, namely the operators
\[
\has \left(E_\al+E'_\al\right)=E_n\mtxt{and}
{\im\ov\sqrt{2}}\left(E_\al-E'_\al\right)\sim
 \has (C_{d-2} + \im C_{d-1})=E_{n+1}\;.\]
Their explicit forms can be found in \refs{expl1} and \refs{expl11}.

\end{appendix}




\begin{thebibliography}{10}

\bibitem{history} J. Hermann, Giornale de Letterati D'Italia \textbf{2} (1710) 447;

J. Bernoulli, Histoires de L'Academie Royale des
Sciences avec les M\'emoires de Mathematique et Physique
(1712) 523;

P.S. de Laplace \emph{Trait\'e de m\'ecanique celeste} Paris An VII. 1798-1799;

C. Runge, \emph{Vektoranalysis, Vol. 1}, Hirzel, Leipzig 1919;

W. Lenz, Z. Phys. \textbf{24} (1924) 197.

\bibitem{goldstein} H. Goldstein, Am. J. Phys. \textbf{43} (1975) 737 and
Am. J. Phys. \textbf{44} (1976) 1123.

\bibitem{pauli} W. Pauli, Z. Phys. \textbf{36} (1926) 336;

L. Hulth\'en, Z. Phys. \textbf{86} (1933) 21;

V. Bargmann, Z. Phys. \textbf{99} (1936) 576.

\bibitem{fock} V. Fock, Z. Phys. \textbf{98} (1935) 145.

\bibitem{zwanziger} D. Zwanziger, Journ. Math. Phys. \textbf{8} (1967) 1858.

\bibitem{sudarshan} E.C.G. Sudarshan, N. Mukunda and L. O'Raifeartaigh, Phys. Lett. \textbf{19} (1965) 322

\bibitem{alliluev} S.P. Alliluev, Soviet Phys. - JETP \textbf{6} (1958) 156.

\bibitem{bander} M. Bander and C. Itzykson, Rev. Mod. Phys. \textbf{38} (1966) 330.

\bibitem{cornwell} J.F. Cornwell, \emph{Group Theory in Physics, Vol. 2}, 
Academic Press, London 1984.

\bibitem{witten1} E. Witten, J. Diff. Geom. \textbf{17} (1982) 661.

\bibitem{witten2} E. Witten, Nucl. Phys. \textbf{B 188} (1981) 513.

\bibitem{nicolai} H. Nicolai, J. Phys. \textbf{A 9} (1976) 1497.

\bibitem{cowi} F. Cooper, A. Khare, R. Musto and A. Wipf, Annals Phys.
\textbf{187} (1988) 1.

\bibitem{andrianov} A.A. Andrianov, N.V. Borisov and M.V. Ioffe, Phys. Lett.
\textbf{A 105} (1984) 19.

\bibitem{mckay} W.G. McKay and J. Patera, \textit{Tables of Dimensions, Indices and Branching Rules
for Representations of Simple Lie Algebras}, Lecture Notes in Pure and Appl. Math. 69, Dekker, New
York 1981.

\bibitem{heumann} R. Heumann, J. Phys. \textbf{A 35} (2002) 7437.

\bibitem{hamermesh} M. Hamermesh, \textit{Group Theory and its Application to Physical Problems},
Dover Publications Inc., New York 1962.

\bibitem{humphreys} J.E. Humphreys, \textit{Introduction to Lie Algebras and Representation Theory},
Springer, New York 1972.

\bibitem{varadarajan} V.S. Varadarajan, \textit{Lie Groups, Lie Algebras and Their Representations},
Graduate Texts in Mathematics 102, Springer, New York 1984.

\bibitem{lie} M. van Leeuwen, A. Cohen and B. Lisser,\textit{LiE 2.1 Manual},
Computer Algebra Group of CWI, Amsterdam.

\bibitem{holten} G.W. Gibbons, R.H. Rietdijk and J.W. van Holten, Nucl. Phys.
\textbf{B 404} (1993) 42.

\end{thebibliography}
\end{document}